\documentclass[preprint,aps,  nofootinbib]{revtex4}
\usepackage{amsmath}
\usepackage{amssymb}
\usepackage{graphicx}
\usepackage{epsf}
\usepackage{amscd}
\usepackage{amsmath}
\usepackage{epsfig}
\usepackage{amssymb}
\usepackage{amsfonts}
\usepackage{amsmath}
\usepackage{graphicx}
\usepackage{bm}

\begin{document}

\title{Leading Chiral Corrections to the Nucleon Generalized Parton
Distributions}
\author{Shung-ichi Ando}
\email{sando@meson.skku.ac.kr}
\affiliation{Department of Physics, Sungkyunkwan University, Suwon 440-746, Korea }
\affiliation{Theory Group, TRIUMF, 4004 Wesbrook Mall, Vancouver, BC V6T 2A3, Canada }
\author{Jiunn-Wei Chen}
\email{jwc@phys.ntu.edu.tw}
\affiliation{Department of Physics and Center for Theoretical Sciences, National Taiwan
University, Taipei 10617, Taiwan }
\author{Chung-Wen Kao}
\email{cwkao@phys.cts.nthu.edu.tw}
\affiliation{Physics Division, National Center for Theoretical Sciences, Hsinchu, Taiwan, }
\affiliation{Department of Physics, Chung Yuan Christian University, Chung-Li 320, Taiwan}
\date{\today }

\begin{abstract}
Using heavy baryon chiral perturbation theory we study the leading chiral
corrections to the complete set of nucleon generalized parton distributions
(GPDs). We compute the leading quark mass and momentum transfer dependence
of the moments of nucleon GPDs through the nucleon off-forward twist-2
matrix elements. These results are then applied to get insight on the GPDs
and their impact parameter space distributions.
\end{abstract}

\maketitle

\section{Introduction}

Recently there have been great interests in generalized parton distributions
(GPDs) of hadrons (see e.g.\ Refs. \cite{GPDs,Ji,R97} for the original work,
and Refs. \cite{GPV,Diehl03,Belitsky:2005qn} for recent reviews). GPDs
relate quite different physical quantities, such as Feynman's parton
distribution functions (PDFs) and hadron form factors, in the same
framework. By generalizing PDFs' one dimensional parton distribution
pictures, GPDs provide the three-dimensional pictures \cite{burkardt,Wigner}%
. Furthermore, GPDs also give information on highly desirable quantities
such as the quark orbital angular momentum contribution to proton spin \cite%
{Ji}.

Useful constraints on the forms of the nucleon GPDs have been obtained at
DESY \cite{DESY} and Jefferson Lab \cite{JLab}. Typically the GPDs
contribute to experimental observables through convolutions; they are not
directly measurable through those experiments. Thus inputs from theories are
important. Valuable insights are gained through the impact parameter
distribution interpretation \cite{burkardt} (see also \cite{Pier}) and model
computations of GPDs [see e.g. \cite{Belitsky:2005qn} for a review].
Recently, lattice techniques have first been applied to compute the moments
of nucleon GPDs\ directly from QCD \cite{QCDSF}. The latest unquenched
lattice results are presented in \cite{LatticeGPD}. However, due to the
limitation of computing power, these calculations often employed $u$ and $d$
quark masses heavier than their physical values and employed big momentum
transfer. Thus, extrapolations in quark mass and momentum transfer are
required to obtain physical results. In general, the quark mass and momentum
transfer dependence have non-analytic structures which should be
incorporated in the parametrization of the extrapolation formulas.
Fortunately, chiral perturbation theory ($\chi $PT) \cite%
{ChPT,GL,HBChPT,ChPTUlf}, which is an\ effective field theory of QCD, can be
applied to extract those non-analytic structures in a systematic,
model-independent way.

Recently $\chi $PT has first been applied to the computation of hadronic
twist-2 matrix elements~\cite{AS,CJ}, which is related to the moments of
PDFs and GPDs through the operator product expansions. Many applications
have been worked out, e.g.,~chiral extrapolations of lattice data \cite%
{AS,CJ,DMNRT,Detmold:2005pt} including (partially) quenching \cite%
{Qtwist2,Beane:2002vq} and finite volume \cite{Detmold:2005pt} effects,
GPDs~for quark contribution to proton spin \cite{JqCJ}, gravitational form
factors \cite{BJ}, pion GPDs \cite{piGPD,piGPD2}, large $N_{C}$ relations
among nucleon and $\Delta $-resonance distributions \cite{Nc} (see also
earlier work in the large $N_{C}$ limit \cite{LargeN}), soft pion
productions in deeply virtual Compton scattering~\cite{DVCSpi,DVCSpi-x,BS},
SU(3) symmetry breaking in the complete set of twist-2 \cite{CS} and twist-3 
\cite{t3} light cone distribution functions, pion-photon transition
distributions \cite{pi-gam} and exclusive semileptonic B decays \cite%
{Bdecays}. The method is also generalized to the multi-nucleon case \cite%
{BS,CW}. There are also earlier results derived using the soft pion theorem 
\cite{Polyakov:1998ze,Polyakov:1999gs,Mankiewicz:1998kg,Penttinen:1999th}.\ 

In this paper, we use $\chi $PT to study the complete set of nucleon GPDs.
We compute the leading quark mass and momentum transfer dependence of the
moments of nucleon GPDs through computing the nucleon off-forward (meaning
the initial and final nucleon momenta being different) twist-2 matrix
elements. Then we apply the $\chi $PT results of the moments to get insight
on the GPDs themselves and their impact parameter space distributions.

\section{Nucleon GPDs}

The eight nucleon GPDs, $H$, $E$, $\tilde{H}$, $\tilde{E}$, $H_{T}$, $\tilde{%
H}_{T}$, $E_{T}$, and $\tilde{E}_{T}$, form the complete set of leading
twist GPDs for quarks of flavor $q$ \cite{Diehl:2001pm}. They are defined
through off-forward nucleon matrix elements of the vector, axial vector and
tensor quark bilinears: 
\begin{eqnarray}
\langle P^{\prime }|\,\overline{q}\left( -\frac{z}{2}n\right) \!\!n\!\!\!%
\slash q\left( \frac{z}{2}n\right) \,\,|P\rangle &=&\int dxe^{-ixzn\cdot 
\bar{P}}\overline{U}(P^{\prime })\left[ Hn\!\!\!\slash+E\,\frac{i\sigma
^{\alpha \beta }n_{\alpha }\Delta _{\beta }}{2M}\right] U(P)\ ,
\label{def_1} \\
\langle P^{\prime }|\,\overline{q}\left( -\frac{z}{2}n\right) n\!\!\!\slash%
\gamma _{5}q\left( \frac{z}{2}n\right) \,\,|P\rangle &=&\int dxe^{-ixzn\cdot 
\bar{P}}\overline{U}(P^{\prime })\left[ \tilde{H}n\!\!\!\slash\gamma _{5}+%
\tilde{E}\,\frac{\gamma _{5}n\cdot \Delta }{2M}\right] U(P)\ ,  \label{def_2}
\end{eqnarray}%
\begin{eqnarray}
\lefteqn{\langle P^{\prime }|\,\overline{q}\left( -\frac{z}{2}n\right) i\
\!n_{\alpha }r_{\beta }\sigma ^{\alpha \beta }q\left( \frac{z}{2}n\right)
\,\,|P\rangle }  \notag \\
&=&\int dxe^{-ixzn\cdot \bar{P}}\overline{U}(P^{\prime })\left[
H_{T}\,in_{\alpha }r_{\beta }\sigma ^{\alpha \beta }+\tilde{H}_{T}\,\frac{%
\left( n\cdot \overline{P}\right) \left( r\cdot \Delta \right) -\left(
r\cdot \overline{P}\right) \left( n\cdot \Delta \right) \,}{M^{2}}\right. 
\notag \\
&&\left. +E_{T}\,\frac{n\!\!\!\slash\left( r\cdot \Delta \right) -r\!\!\!%
\slash\left( n\cdot \Delta \right) \,}{2M}+\tilde{E}_{T}\,\frac{n\!\!\!\slash%
\left( r\cdot \overline{P}\right) -r\!\!\!\slash\left( n\cdot \overline{P}%
\right) \,}{2M}\right] U(P)\ ,  \label{def_3}
\end{eqnarray}%
where $M$ is the nucleon mass, $U$ is the nucleon Dirac spinor with
normalization $\overline{U}(P)U(P)=2M$, $\bar{P}^{\mu }=(P+P^{\prime })^{\mu
}/2$, $\Delta ^{\mu }=P^{\mu \prime }-P^{\mu }$, $n$ is a dimensionless
light-like vector [$n^{2}=0$], $\xi =-n\cdot \Delta /\left( 2n\cdot \bar{P}%
\right) $ and $t=\Delta ^{2}$. $r$ is a transverse vector satisfying $r\cdot
n=r\cdot \bar{P}=0$. We have used $\epsilon ^{0123}=-1$ and the notation $%
\epsilon ^{ABCD}=\epsilon ^{\alpha \beta \gamma \delta }A_{\alpha }B_{\beta
}C_{\gamma }D_{\delta }$. Here we have suppressed the Wilson lines
connecting quark fields locating at $-\frac{z}{2}n$ and $\frac{z}{2}n$ to
make the nonlocal quark operators gauge invariant. These GPDs are functions
of $x$, $\xi $, and $t$. Both $x$ and $\xi $ have support from $-1$ to $+1$.

The GPDs encode information of ordinary PDFs and nucleon form factors \cite%
{Ji}. In the forward limit of $\Delta ^{\mu }\rightarrow 0$, we have 
\begin{equation}
H(x,0,0)=f_{1}(x),~~~\tilde{H}(x,0,0)=g_{1}(x),\quad H_{T}(x,0,0)=f_{T}(x),
\label{f1g1}
\end{equation}%
where $f_{1}(x)$, $g_{1}(x)$ and $f_{T}(x)$ are spin-averaged, helicity and
transversity PDFs, respectively. On the other hand, forming the first $x$
moments of the new distributions, one gets the following sum rules, 
\begin{eqnarray}
\int dxH(x,\xi ,t) &=&F_{1}(t)\ ,  \notag \\
\int dxE(x,\xi ,t) &=&F_{2}(t)\ ,  \notag \\
\int dx\tilde{H}(x,\xi ,t) &=&G_{A}(t)\ ,  \notag \\
\int dx\tilde{E}(x,\xi ,t) &=&G_{P}(t)\ .  \label{F1F2}
\end{eqnarray}%
where $F_{1}$ and $F_{2}$ are the Dirac and Pauli form factors and $G_{A}$
and $G_{P}$ are the axial-vector and pseudo-scalar form factors. There are
also tensor form factors associated with the first $x$ moments of $H_{T}$, $%
\tilde{H}_{T}$ and $E_{T}$, but not $\tilde{E}_{T}$,\ because time reversal
invariance demands \cite{Diehl:2001pm} 
\begin{equation}
\int dx\tilde{E}_{T}(x,\xi ,t)=0\ .
\end{equation}

The most interesting sum rule relevant to the nucleon spin is \cite{Ji}, 
\begin{equation}
J_{q}=\frac{1}{2}\int dxx[H(x,\xi ,0)+E(x,\xi ,0)]\ ,  \label{eq1}
\end{equation}%
where $J_{q}$ is the $q$ quark contribution to proton spin in a frame in
which the proton has a definite helicity. The $\xi $ dependence in the sum
rule has dropped out. Since $J_{q}$ can further be decomposed into quark
helicity and orbital angular moment contributions, by measuring $J_{q}$ from
experiments sensitive to GPDs and measuring the helicity contribution from
polarized deep inelastic scattering, the quark orbital angular momentum
contribution to proton spin in principle can be obtained.

There are also a set of eight gluon GPDs defined as the matrix elements of
non-local gluon operators. They mix with the isoscalar combination of quark
GPDs under renormalization scale and transform in the same way as isoscalar
quark GPDs under chiral transformation. We will first focus on the quark
GPDs, and later come back to the gluon GPDs.

\section{Chiral Perturbation Theory}

$\chi $PT is a low-energy effective field theory of QCD. $\chi $PT makes use
of the symmetries and scale separation of QCD and allows a model independent
description of physics below the chiral symmetry breaking scale $\mu _{\chi
}(\sim 4\pi F_{\pi }\sim 1$ GeV$)$, where $F_{\pi }=93$ MeV is the pion
decay constant. In this paper, we will compute single nucleon matrix
elements associated with nucleon GPDs; thus, the relevant scales below $\mu
_{\chi }$ (light scales) include the pion mass $m_{\pi }\simeq 139$ MeV and
the characteristic momentum $p$ in the problem. The nucleon mass $M$, which
is numerically of the same size as $\mu _{\chi }$, is treated as a heavy
scale as $\mu _{\chi }$.\ Thus the standard heavy baryon $\chi $PT \cite%
{HBChPT}\ approach is used to systematically disentangle the light and heavy
scales. Here the following four small expansion parameters are treated the
same in the chiral expansion and denoted as 
\begin{equation}
\varepsilon =\frac{p}{\mu _{\chi }},\frac{m_{\pi }}{\mu _{\chi }},\frac{p}{M}%
,\frac{m_{\pi }}{M}.  \label{epsilon}
\end{equation}

The physical pion fields ($\pi ^{0}$, $\pi ^{+}$, $\pi ^{-}$) enter the
theory through the matrices 
\begin{equation}
\Sigma =e^{i\Pi /F_{\pi }}\,,\quad \quad u=\sqrt{\Sigma }\,,
\end{equation}%
where 
\begin{equation}
\Pi =%
\begin{pmatrix}
\pi ^{0} & \sqrt{2}\pi ^{+} \\ 
\sqrt{2}\pi ^{-} & -\pi ^{0}%
\end{pmatrix}%
\,.
\end{equation}%
The relevant terms in the chiral Lagrangian are%
\begin{eqnarray}
\mathcal{L} &=&{\frac{F_{\pi }^{2}}{8}}\mathrm{Tr}\left[ \ \partial ^{\mu
}\Sigma \ \partial _{\mu }\Sigma ^{\dagger }\ \right] \ +\ \lambda \ \mathrm{%
Tr}\left[ \ m_{q}\Sigma ^{\dagger }\ +\ \mathrm{h.c.}\ \right] +  \notag \\
&&\mathcal{+}iN^{\dagger }v\cdot DN+\ 2g_{A}N^{\dagger }\ S\cdot \mathcal{A}%
\ N+\cdots \,,  \label{L_1}
\end{eqnarray}%
where the quark mass matrix $m_{q}=diag(m_{u},m_{d})=m_{q}^{\dag }$, and we
will take the isospin symmetry limit $m_{u}=m_{d}=\overline{m}$. The nucleon
field $N=(p,n)^{T}$, $v$ is the nucleon velocity and $S^{\mu }=\frac{i}{2}%
\sigma ^{\mu \nu }\gamma _{5}v_{\nu }$ is the nucleon spin vector. $v\cdot
S=0$. $g_{A}=1.26$ is the axial-vector coupling constant. The pion-nucleon
couplings arise in Eq.~(\ref{L_1}) through the vector and axial couplings 
\begin{equation}
\mathcal{V}^{\mu }=\frac{1}{2}\left( u\partial ^{\mu }u^{\dag }+u^{\dag
}\partial ^{\mu }u\right) \,,\qquad \qquad \mathcal{A}^{\mu }=\frac{i}{2}%
\left( u\partial ^{\mu }u^{\dagger }-u^{\dagger }\partial ^{\mu }u\right) \,,
\end{equation}%
and the chiral covariant derivative 
\begin{equation}
D^{\mu }=\left( \partial ^{\mu }+\mathcal{V}^{\mu }\right) \,.
\end{equation}%
Under a $SU(2)_{L}\otimes SU(2)_{R}$ chiral rotation, the hadronic fields in
the chiral lagrangian transform as 
\begin{gather}
\Sigma \rightarrow L\Sigma R^{\dag }\ ,\quad m_{q}\rightarrow Lm_{q}R^{\dag
}\ ,  \notag \\
N\rightarrow \mathcal{U}(x)N\ ,\quad u\rightarrow Lu\mathcal{U}^{\dagger
}(x)=\mathcal{U}(x)uR^{\dagger }\ ,  \notag \\
\mathcal{A}^{\mu }\rightarrow \mathcal{U}(x)\mathcal{A}^{\mu }\mathcal{U}%
^{\dagger }(x)\ ,\quad D^{\mu }N\rightarrow \mathcal{U}(x)D^{\mu }N\ ,
\end{gather}%
such that the chiral lagrangian in Eq.~(\ref{L_1}) stays invariant under
chiral transformation.

\section{The Vector Operators}

Instead of directly matching the non-local quark bilinear operators to
hadronic operators in $\chi $PT, it is conceptually more straightforward to
deal with matching of local operators. One can perform operator product
expansions (OPEs) to convert the non-local operators to the sums of local
twist-2 operators then do the matching \cite{AS,CJ}. The Taylor-series
expansion of Eq.(\ref{def_1}) gives%
\begin{equation}
\langle P^{\prime }|\mathcal{O}^{m}|P\rangle =\ \left( n\cdot \bar{P}\right)
^{m}\overline{U}(P^{\prime })\left[ H_{m+1}n\!\!\!\slash+E_{m+1}\frac{%
i\sigma ^{\alpha \beta }n_{\alpha }\Delta _{\beta }}{2M}\right] U(P)\ ,
\label{a0}
\end{equation}%
where 
\begin{equation}
\mathcal{O}^{m}=\overline{q}n\!\!\!\slash\left( in\cdot \tensor{D}\right)
^{m}q
\end{equation}%
is a twist-2 operator dotted by the $n^{\mu _{0}}n^{\mu _{1}}\cdots n^{\mu
_{n}}$ tensor to project out the symmetric and traceless part. The gauge
invariant covariant derivative $\tensor{D}^{\mu }=(\overrightarrow{D}^{\mu }-%
\overleftarrow{D}^{\mu })/2$. $H_{m+1}(\xi ,t)=\int dxx^{m}H(x,\xi ,t)\ $and 
$E_{m+1}(\xi ,t)=\int dxx^{m}E\,(x,\xi ,t)$ are the $\left( m+1\right) $-th
moments in $x$ of the GPDs. The nucleon matrix element of $O^{m}$ has
different form factor structures \cite{Ji,Ji:2000id,Hagler:2004yt}. Using
the notation of Ref. \cite{Hagler:2004yt}, 
\begin{eqnarray}
\langle P^{\prime }|\mathcal{O}^{m}|P\rangle &=&\overline{U}(P^{\prime })%
\left[ \sum_{\substack{ j=0  \\ even}}^{m}\left\{ n\!\!\!\slash\left( n\cdot
\Delta \right) ^{j}\left( n\cdot \bar{P}\right) ^{m-j}A_{m+1,j}(t)\right.
\right.  \notag \\
&&\left. -i\frac{\Delta _{\alpha }n_{\mu }\sigma ^{\alpha \mu }}{2M}\left(
n\cdot \Delta \right) ^{j}\left( n\cdot \bar{P}\right)
^{m-j}B_{m+1,j}(t)\right\}  \notag \\
&+&\left. \left. \frac{1}{M}\left( n\cdot \Delta \right)
^{m+1}C_{m+1}(t)\right\vert _{m\ odd}\right] U(P)\ .  \label{eq: relvector}
\end{eqnarray}%
The constraints on $j$ and $C_{m}$ are due to the requirement of time
reversal invariance \cite{Ji}. By comparing this equation with Eq.(\ref{a0}%
), we have 
\begin{eqnarray}
H_{m+1}(\xi ,t) &=&\int_{-1}^{1}dxx^{m}H(x,\xi ,t)=\sum_{\substack{ j=0  \\ %
even}}^{m}(-2\xi )^{j}~A_{m+1,j}(t)+(-2\xi )^{m+1}\left.
C_{m+1}(t)\right\vert _{m\ odd}\ ,  \notag \\
E_{m+1}(\xi ,t) &=&\int_{-1}^{1}dxx^{m}E(x,\xi ,t)=\sum_{\substack{ j=0  \\ %
even}}^{m}(-2\xi )^{j}~B_{m+1,j}(t)-(-2\xi )^{m+1}\left.
C_{m+1}(t)\right\vert _{m\ odd}\ ,  \label{aa}
\end{eqnarray}%
after using the Gordon decomposition. The time reversal invariance demands $%
H_{m+1}$ and $E_{m+1}$ to be even in $\xi $:%
\begin{equation}
H_{m+1}(-\xi ,t)=H_{m+1}(\xi ,t)\ ,\quad E_{m+1}(-\xi ,t)=E_{m+1}(\xi ,t)\ .
\end{equation}

To apply heavy baryon $\chi $PT, we perform the $1/M$ expansion to Eq.(\ref%
{eq: relvector}) \cite{HBChPT}. The leading terms are 
\begin{eqnarray}
\langle P^{\prime }|\mathcal{O}^{m}|P\rangle &=&2\overline{N}\left[ \sum 
_{\substack{ j=0  \\ even}}^{m}\left\{ \left( n\cdot \Delta \right)
^{j}\left( Mn\cdot v\right) ^{m-j+1}E_{m+1,j}(t)\right. \right.  \notag \\
&&+\left. i\epsilon ^{vn\Delta S}\,\left( n\cdot \Delta \right) ^{j}\left(
Mn\cdot v\right) ^{m-j}M_{m+1,j}(t)\right\}  \notag \\
&&+\left. \left. \left( n\cdot \Delta \right) ^{m+1}C_{m+1}(t)\right\vert
_{m\ odd}\right] N\ ,  \label{VecOp}
\end{eqnarray}%
where 
\begin{equation}
E_{m+1,j}=A_{m+1,j}+\frac{t}{4M^{2}}B_{m+1,j}\ ,\quad
M_{m+1,j}=A_{m+1,j}+B_{m+1,j}\ .  \label{xx}
\end{equation}%
For convenience, we will work in the Breit frame where $v^{\mu }=(1,\vec{0})=%
\bar{P}^{\mu }/M+\mathcal{O}(1/M)$, $S^{\mu }=(0,\vec{\sigma}/2)$ and $%
\Delta ^{\mu }=(0,\vec{\Delta}~)$. The normalization of the Pauli spinor $N$
is $\overline{N}N=1+\mathcal{O}(1/M)$.

\subsection{Pionic Vector Operators}

In a similar manner, the pion vector GPD is defined as 
\begin{eqnarray}
&&\langle \pi ^{i}(P_{\pi }^{\prime })|\,\overline{q}\left( -\frac{z}{2}%
n\right) \tau ^{\alpha }\!\!\ n\!\!\!\slash q\left( \frac{z}{2}n\right) |\pi
^{j}(P_{\pi })\rangle  \notag \\
&=&\int dye^{-iyzn\cdot \bar{P}_{\pi }}H_{\pi }^{\alpha }(y,\xi _{\pi
},t)n\cdot \bar{P}_{\pi }\text{tr}\left[ \ \!\!\tau ^{i}\tau ^{\alpha }\tau
^{j}\right] \ ,
\end{eqnarray}%
where the isospin operator $\tau ^{a}=\left( 1,\overrightarrow{\tau }\right) 
$, and from now on $q$ is an isospin multiplet [but note that the isoscalar
quark also contains the $s$ quark contribution]. There is no $E_{\pi }$ GPD
because the pion is spinless. The Taylor-series expansion of the above
equation gives 
\begin{equation}
\langle \pi ^{i}(P_{\pi }^{\prime })|\mathcal{O}_{\alpha }^{m}|\pi
^{j}(P_{\pi })\rangle =H_{\pi ,m+1}^{\alpha }(\xi _{\pi },t)\left( n\cdot 
\bar{P}_{\pi }\right) ^{m+1}tr\left[ \ \!\!\tau ^{i}\tau ^{\alpha }\tau ^{j}%
\right] \ ,  \label{H_pi}
\end{equation}%
where $\bar{P}_{\pi }=\left( P_{\pi }+P_{\pi }^{\prime }\right) /2$ and 
\begin{equation}
\mathcal{O}_{\alpha }^{m}=\bar{q}\ \tau ^{\alpha }n\!\!\!\slash(n\cdot i%
\tensor{D})^{m}q\ ,
\end{equation}%
and $H_{\pi ,m+1}(\xi _{\pi },t)=\int dyy^{m}H_{\pi }(y,\xi _{\pi },t)$.

The pion GPDs are strongly constrained by charge conjugation ($C$) and
isospin symmetry. Under $C$,%
\begin{equation}
\mathcal{CO}_{\alpha }^{m}\mathcal{C}^{-1}=\left( -1\right) ^{m+1}\mathcal{O}%
_{\alpha }^{m}\ ,  \label{C}
\end{equation}%
for $\alpha =0$ and $3$ \cite{CS}. For $m$ even ($m=2k$), the above equation
implies

\begin{equation}
\left\langle \pi ^{0}(P_{\pi }^{\prime })|\mathcal{O}_{0}^{2k}|\pi
^{0}(P_{\pi })\right\rangle =0\ ,\quad \left\langle \pi ^{0}(P_{\pi
}^{\prime })|\mathcal{O}_{3}^{2k}|\pi ^{0}(P_{\pi })\right\rangle =0\ ,
\end{equation}%
because $\pi ^{0}$ is $C$ even. Furthermore, by isospin symmetry, $\langle
\pi ^{\pm }(P_{\pi }^{\prime })|O_{0}^{2k}|\pi ^{\pm }(P_{\pi })\rangle =0$.
This implies 
\begin{equation}
H_{\pi ,2k+1}^{0}(\xi _{\pi },t)=0\ ,
\end{equation}%
or, equivalently,%
\begin{equation}
H_{\pi }^{0}(y,\xi _{\pi },t)=H_{\pi }^{0}(-y,\xi _{\pi },t)\ .
\end{equation}

For $m$ odd ($m=2k-1$),%
\begin{eqnarray}
&&\langle \pi ^{i}(P_{\pi }^{\prime })|\mathcal{O}_{0}^{2k-1}|\pi
^{j}(P_{\pi })\rangle  \notag \\
&=&2\delta _{ij}\left\{ \sum_{l=0}^{k-1}\left( n\cdot \Delta \right)
^{2l}\left( n\cdot \bar{P}_{\pi }\right) ^{2k-2l}A_{2k,2l}^{\pi
,0}(t)+\left( n\cdot \Delta \right) ^{2k}C_{2k}^{\pi ,0}(t)\right\} \ ,
\label{pi_1}
\end{eqnarray}%
where $P_{\pi }^{\prime }-P_{\pi }=\Delta $ and $\xi _{\pi }=-n\cdot \Delta
/\left( 2n\cdot \bar{P}_{\pi }\right) $.

On the other hand, for the isovector case, under $C$ 
\begin{eqnarray}
\langle \pi ^{+}(P_{\pi }^{\prime })|O_{3}^{2k-1}|\pi ^{+}(P_{\pi })\rangle
&\rightarrow &\langle \pi ^{-}(P_{\pi }^{\prime })|O_{3}^{2k-1}|\pi
^{-}(P_{\pi })\rangle \   \notag \\
&=&-\langle \pi ^{+}(P_{\pi }^{\prime })|O_{3}^{2k-1}|\pi ^{+}(P_{\pi
})\rangle \ ,
\end{eqnarray}%
where the equals sign is due to $\langle O_{3}^{2k-1}\rangle \propto \langle
\tau _{3}\rangle $. This, together with $\langle \pi ^{0}|O_{3}^{2k-1}|\pi
^{0}\rangle =0$, implies%
\begin{equation}
H_{\pi ,2k}^{3}(\xi _{\pi },t)=0\ .
\end{equation}%
Thus

\begin{equation}
\langle \pi ^{i}(P_{\pi }^{\prime })|O_{3}^{2k}|\pi ^{j}(P_{\pi })\rangle
=2i\epsilon ^{i3j}\sum_{l=0}^{k}\left( n\cdot \Delta \right) ^{2l}\left(
n\cdot \bar{P}_{\pi }\right) ^{2k-2l+1}A_{2k+1,2l}^{\pi ,3}(t)\ .
\label{pi_2}
\end{equation}

Again, time reversal invariance requires $H_{\pi }^{\alpha }(y,\xi _{\pi
},t) $ to be even in $\xi _{\pi }$: 
\begin{equation}
H_{\pi }^{\alpha }(y,\xi _{\pi },t)=H_{\pi }^{\alpha }(y,-\xi _{\pi },t)\ ,
\end{equation}%
as shown in Eqs. (\ref{pi_1}) and (\ref{pi_2}).

\bigskip In matching $O_{\alpha }^{m}$\ to the hadronic operators in $\chi $%
PT, it is useful to write 
\begin{equation}
\mathcal{O}_{\alpha }^{m}=\mathcal{O}_{\alpha ,R}^{m}+\mathcal{O}_{\alpha
,L}^{m}\ ,  \label{vector}
\end{equation}%
where 
\begin{equation}
\mathcal{O}_{\alpha ,R}^{m}=\bar{q}_{R}\ \tau _{R}^{\alpha }n\!\!\!\slash%
(n\cdot i\tensor{D})^{m}q_{R}\ ,
\end{equation}%
and similarly for $O_{\alpha ,L}^{m}$. $q_{L,R}=[(1\mp \gamma _{5})/2]q$ is
the left(right)-handed quark field. The distinction between $\tau _{L}^{a}$
and $\tau _{R}^{a}$ is only for bookkeeping purposes. We will set $\tau
_{L}^{a}=\tau _{R}^{a}=\tau ^{a}$ at the end. Under a global chiral $%
SU(2)_{L}\times SU(2)_{R}$ transformation, $q_{R}\rightarrow Rq_{R}$ and $%
q_{L}\rightarrow Lq_{L}$. If we demand

\begin{equation}
\tau _{L}^{a}\rightarrow L\tau _{L}^{a}L^{\dag },\qquad \tau
_{R}^{a}\rightarrow R\tau _{R}^{a}R^{\dag },  \label{pi_t}
\end{equation}%
then $O_{\alpha ,R(L)}^{m}$ will be invariant under chiral transformation.
Furthermore, under charge conjugation, $\Sigma \rightarrow \Sigma ^{T}$, $%
m_{q}\rightarrow m_{q}^{T}$, and if we demand 
\begin{equation}
\tau _{L}^{a}\rightarrow \tau _{L}^{a\,T}\,,\qquad \tau _{R}^{a}\rightarrow
\tau _{R}^{a\,T},  \label{lambda}
\end{equation}%
then Eq.(\ref{C}) can be satisfied for any $\alpha $.

Using the symmetries mentioned above, we now match $O_{\alpha }^{m}$\ to the
most general combination of hadronic operators with the same symmetries, 
\begin{equation}
\mathcal{O}_{\alpha }^{m}\,\rightarrow O_{\alpha ,\pi }^{m}+O_{\alpha
,N}^{m}+\cdots \ ,
\end{equation}%
where $O_{\alpha ,\pi }^{m}$ denotes hadronic operators made of purely pion
fields and $O_{\alpha ,N}^{m}$ denotes hadronic operators with nucleon
number equal to one. The ellipse denotes operators which do not contribute
to our SU(2) calculations in single nucleon sectors, such as operators with
nucleon number equal to two and above or operators with hyperon fields.

For a given $m,$the leading pionic operators in the chiral expansion are

\begin{eqnarray}
O_{\alpha ,\pi }^{m} &=&\frac{F_{\pi }^{2}}{4}\sum_{\substack{ j=0  \\ even}}%
^{m}\overline{A}_{m+1,j}^{\pi ,\alpha }(0)\left( -in\cdot \partial \right)
^{j}tr\left[ \tau _{L}^{\alpha }\!\!\ \Sigma \left( in\cdot \tensor{\partial}%
\right) ^{m-j+1}\Sigma ^{\dagger }\right.  \notag \\
&&\left. +\tau _{R}^{\alpha }\!\!\ \Sigma ^{\dagger }\left( in\cdot %
\tensor{\partial}\right) ^{m-j+1}\Sigma \right] +\cdots \ ,  \label{pi-mat}
\end{eqnarray}%
where $\tensor{\partial}^{\mu }=(\overrightarrow{\partial }^{\mu }-%
\overleftarrow{\partial }^{\mu })/2$ and the ellipse denotes higher order
operators with more powers of derivatives or quark mass matrix. $m=0,1,2...$
There is no restriction on the value of $m$ \cite{AS,CJ}. For $\alpha =0$
(the isoscalar case), it is easy to see that the contribution with $m$ even
vanishes---a consequence of charge conjugation. Using 
\begin{eqnarray*}
\left( -in\cdot \partial \right) ^{2m}\left[ \Sigma \left( in\cdot %
\tensor{\partial}\right) ^{2n}\Sigma ^{\dagger }\right] &=&\left( -in\cdot
\partial \right) ^{2m+2}\left[ \Sigma \left( in\cdot \tensor{\partial}%
\right) ^{2n-2}\Sigma ^{\dagger }\right] \\
&&-\left( -in\cdot \partial \right) ^{2m}\left[ \left( in\cdot \partial
\Sigma \right) \left( in\cdot \tensor{\partial}\right) ^{2n-2}\left( in\cdot
\partial \Sigma ^{\dagger }\right) \right] \ ,
\end{eqnarray*}%
the operator in Eq.(\ref{pi-mat}) can be rewritten as%
\begin{equation}
O_{0,\pi }^{2k-1}\,=\frac{F_{\pi }^{2}}{2}\sum_{l=0}^{k-1}A_{2k,2l}^{\pi
,0}(0)\left( -in\cdot \partial \right) ^{2l}tr\left[ \left( in\cdot \partial
\Sigma \right) \left( in\cdot \tensor{\partial}\right) ^{2k-2l-2}\left(
in\cdot \partial \Sigma ^{\dagger }\right) \right] +\mathcal{\cdots }\ ,
\label{m1}
\end{equation}%
which is the operator constructed in Ref. \cite{piGPD}. The prefactors in
Eq.(\ref{m1}) are chosen such that Eq.(\ref{pi_1}) is reproduced in the
leading order in the chiral expansion subject to constraints that relate $%
C_{2k}^{\pi ,0}$ to $A_{2k,2l}^{\pi ,0}$. The constraints come from the $%
\left( n\cdot P_{\pi }\right) \left( n\cdot P_{\pi }^{\prime }\right)
=\left( n\cdot \bar{P}_{\pi }\right) ^{2}\left( 1-\xi _{\pi }^{2}\right) $
factor in $\langle \pi ^{i}(P_{\pi }^{\prime })|O_{0}^{2k-1}|\pi ^{j}(P_{\pi
})\rangle $ which makes $H_{\pi }^{0}(y,\xi _{\pi }^{2}=\pm 1,0)=0$ [this
was also observed in Refs. \cite{Polyakov:1998ze,Polyakov:1999gs}]. This
property, however, does not persist at higher orders. At the next-to-leading
order (NLO), there several sets of counterterms: 
\begin{eqnarray}
O_{0,\pi ,1}^{2k-1} &=&a_{2k}^{\pi ,0}\left( -in\cdot \partial \right)
^{2k}tr\left[ \Sigma m_{q}^{\dagger }+m_{q}\Sigma ^{\dagger }\right] \ , 
\notag \\
O_{0,\pi ,2}^{2k-1} &=&\sum_{l=0}^{k-1}b_{2k,2l}^{\pi ,0}tr\left[ \Sigma
m_{q}^{\dagger }+m_{q}\Sigma ^{\dagger }\right] \left( -in\cdot \partial
\right) ^{2l}tr\left[ \left( in\cdot \partial \Sigma \right) \left( in\cdot %
\tensor{\partial}\right) ^{2k-2l-2}\left( in\cdot \partial \Sigma ^{\dagger
}\right) \right] \ ,  \notag \\
O_{0,\pi ,3}^{2k-1} &=&\sum_{l=0}^{k-1}c_{2k,2l}^{\pi ,0}\left( -in\cdot
\partial \right) ^{2l}tr\left[ \left( \Sigma m_{q}^{\dagger }+m_{q}\Sigma
^{\dagger }\right) \left( \left( in\cdot \partial \Sigma \right) \left(
in\cdot \tensor{\partial}\right) ^{2k-2l-2}\left( in\cdot \partial \Sigma
^{\dagger }\right) \right) \right] \ ,  \notag \\
O_{0,\pi ,4}^{2k-1} &=&\sum_{l=0}^{k-1}d_{2k,2l}^{\pi ,0}\partial ^{2}\left(
-in\cdot \partial \right) ^{2l}tr\left[ \left( in\cdot \partial \Sigma
\right) \left( in\cdot \tensor{\partial}\right) ^{2k-2l-2}\left( in\cdot
\partial \Sigma ^{\dagger }\right) \right] \ .
\end{eqnarray}%
It is $O_{0,\pi ,1}^{2k-1}$ that makes $H_{\pi }^{0}(y,\pm 1,0)$
non-vanishing.

\subsection{ Nucleon Vector Operators}

In matching $O_{\alpha }^{m}$ to nucleon operators, it is convenient to
define 
\begin{equation}
\tau _{u\pm }^{a}=\frac{1}{2}\left( u\tau _{R}^{a}u^{\dagger }\pm u^{\dagger
}\tau _{L}^{a}u\right) \ ,
\end{equation}%
such that under a chiral transformation, $\tau _{u\pm }^{a}\rightarrow 
\mathcal{U}(x)\tau _{u\pm }^{a}\mathcal{U}(x)^{\dagger }$ with $\tau
_{R(L)}^{a}$ transforms as in Eq.(\ref{pi_t}). The leading operators
contributing to $H_{m+1}^{\alpha }$ and $E_{m+1}^{\alpha }$ are 
\begin{align}
O_{\alpha ,N}^{m}& =2M\left( Mn\cdot v\right) ^{m}\overline{N}\left[
H_{m+1}^{\alpha (0)}(\xi ,0)n\cdot v+\overline{E}_{m+1}^{\alpha (0)}(\xi ,0)%
\frac{i\epsilon ^{vn\Delta S}\,}{M}\right] \tau _{u+}^{\alpha }N  \notag \\
& +2M\left( Mn\cdot v\right) ^{m}\overline{N}\left[ C_{m+1}^{\alpha (0)}(\xi
,0)n\cdot S+D_{m+1}^{\alpha (0)}(\xi ,0)\frac{i\epsilon ^{Sn\Delta S}\,}{2M}%
\right] \tau _{u-}^{\alpha }N+\cdots ,  \label{eq2}
\end{align}%
where $\overline{E}_{m+1}=H_{m+1}+E_{m+1}$. The prefactors are chosen such
that Eq.(\ref{a0}) is reproduced in the leading order in the chiral
expansion and a superscript $(0)$ denotes the chiral limit value. The $%
C_{m+1}$ and $D_{m+1}$ operators will not contribute at the order we are
working (NLO). The $H_{m+1}$ and $\overline{E}_{m+1}$ operators in Eq.(\ref%
{eq2}) can be further written as 
\begin{eqnarray}
O_{\alpha ,N}^{m} &=&2\overline{N}\left[ \sum_{\substack{ j=0  \\ even}}%
^{m}\left\{ \left( n\cdot \Delta \right) ^{j}\left( Mn\cdot v\right)
^{m-j+1}E_{m+1,j}^{\alpha (0)}(0)\right. \right.  \notag \\
&&+\left. i\epsilon ^{vn\Delta S}\,\left( n\cdot \Delta \right) ^{j}\left(
Mn\cdot v\right) ^{m-j}M_{m+1,j}^{\alpha (0)}(0)\right\}  \notag \\
&+&\left. \left. \left( n\cdot \Delta \right) ^{m+1}C_{m+1}^{\alpha
(0)}(0)\right\vert _{m\ odd}\right] \tau _{u+}^{\alpha }N+\cdots \ ,
\label{O_N}
\end{eqnarray}%
where we have replaced the total derivative operator $-in\cdot D$ by $n\cdot
\Delta $. The operators with derivatives acting only on $\tau _{u+}^{a}$ or $%
\tau _{u-}^{a}$ do not contribute to nucleon GPDs by direct computation.

\subsection{Form factor results of vector twist-2 matrix elements}

Now we present the leading chiral corrections to the form factors of the
vector twist-2 operators defined in Eq.(\ref{VecOp}). We will insert powers
of $\varepsilon $ to keep track of the chiral expansion. One should set $%
\varepsilon =1$ when using these results. The results of $E_{m+1,0}^{\alpha
}(0)$ reproduce those of the forward twist-2 matrix elements in Refs. \cite%
{AS,CJ,Nc}, while $E_{1,0}^{\alpha }(t)$ and $M_{1,0}^{\alpha }(t)$ agree
with the electric and magnetic form factor results in $\chi $PT \cite%
{ChPTUlf}.

\begin{figure}[t]
\vskip 0.cm 
\centerline{
  \mbox{\epsfxsize=14.0truecm \hbox{\epsfbox{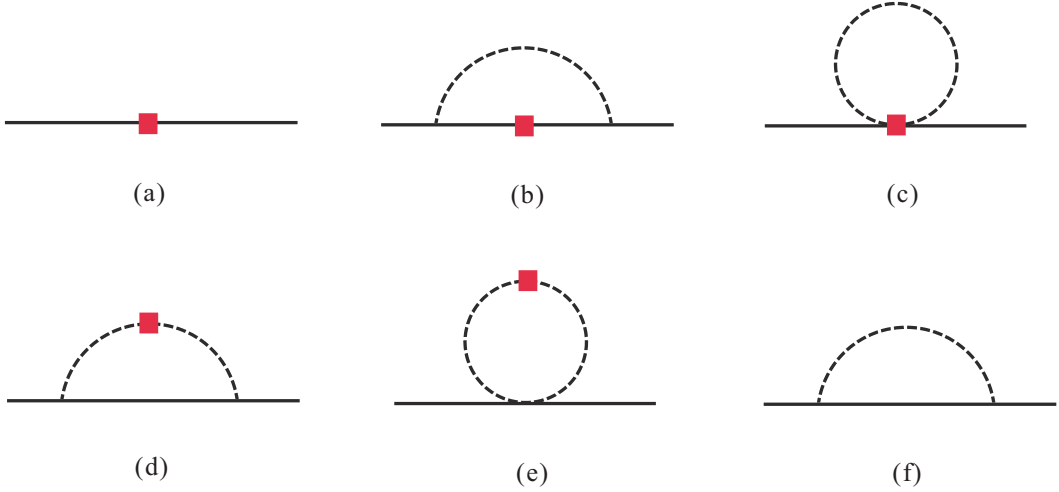}} }
  } \vskip-0.1cm \vspace{-0.5cm} 
\caption[1]{Diagrams contributing\ to\ leading chiral corrections of nucleon
off-forward twist-2 matrix elements.\ The squares denote insertions of
hadronic operators. The solid lines are nucleon fields and the dashed lines
are pion fields. The diagram (f) denotes the wave function renormalization
contribution. The leading order pion pole diagram is shown in Fig.2.}
\label{Fig1}
\end{figure}

For the isoscalar ($\alpha =0$) form factors\footnote{%
There is no restriction on the range of $m$ in this ChPT calculation. We are
only interested in the $m_{\pi }$ and $t$ dependence of the form factors,
thus we keep track of the expansion in $m_{\pi }$ and $t$ using the
parameter $\varepsilon $ and count $x=O(\varepsilon ^{0})$. There is no
intrinsic difference between operators of different $m$---they all transform
in the same way under a chiral rotation \cite{AS,CJ}.},

\begin{eqnarray}
E_{m+1,2k}^{0}(t) &=&E_{m+1,2k}^{0(0)}(0)+\varepsilon ^{2}\left( 1-\delta
_{m,0}\delta _{k,0}\right) \alpha _{m+1,2k}^{0}m_{\pi }^{2}+\varepsilon
^{2}\beta _{m+1,2k}^{0}t+\mathcal{O}(\varepsilon ^{3})\ ,  \notag \\
M_{m+1,2k}^{0}(t) &=&\left( 1+\varepsilon ^{2}\mathcal{C}_{M}^{0}\right)
M_{m+1,2k}^{0(0)}(0)+\varepsilon ^{2}\delta _{m,2k+1}\mathcal{D}%
_{2k+2,2k}^{0,M}\left( t\right)  \notag \\
&&+\varepsilon ^{2}\gamma _{m+1,2k}^{0}m_{\pi }^{2}+\varepsilon ^{2}\zeta
_{m+1,2k}^{0}t+\mathcal{O}(\varepsilon ^{3})\ ,  \notag \\
C_{m+1}^{0}(t) &=&C_{m+1}^{0(0)}(0)+\varepsilon \delta _{m,2k+1}\mathcal{D}%
_{2k+2}^{C}\left( t\right) +\mathcal{O}(\varepsilon ^{2})\text{ ,}
\label{r1}
\end{eqnarray}%
where $k=0,1,2\ldots $ The $\mathcal{C}$\ contribution is from the wave
function renormalization (Fig.1(f)) and the loop diagrams with one insertion
of the nucleon operators in Eq. (\ref{O_N}) (Figs. 1(b) and (c)). The $%
\mathcal{D}$\ contributions are from the loop diagrams with one insertion of
the pionic operators in Eq. (\ref{O_N}) (Figs. 1 (d) and (e)).

For the $E_{m+1,2k}^{0}(t)$ form factor, the one-loop contributions from
Figs. 1(b) and (f) cancel each other while the diagrams in Figs. 1(d) and
(e) are of higher order and Fig. 1(c) vanishes. At $\mathcal{O}(\varepsilon
^{2})$, $E_{m+1,2k}^{0}$ also receives analytic contributions proportional
to $m_{\pi }^{2}$ or $t$ from counterterms (Fig. 1(a)) except for the charge
operator $E_{1,0}^{0}(0)$. $\alpha _{m+1,2k}^{0}$ and $\beta _{m+1,2k}^{0}$
are independent of $m_{\pi },$ $t$ and the renormalization scale $\mu $.

For the $M_{m+1,2k}^{0}$ form factor, 
\begin{equation}
\mathcal{C}_{M}^{0}=-\frac{3g_{A}^{2}m_{\pi }^{2}}{(4\pi F_{\pi })^{2}}\ln 
\frac{m_{\pi }^{2}}{\mu ^{2}}\ ,
\end{equation}%
and%
\begin{equation}
\mathcal{D}_{2k+2,2k}^{0,M}\left( t\right) =\frac{3g_{A}^{2}}{16{\pi }%
^{2}F_{\pi }^{2}}\sum_{l=0}^{k}\,A_{2k+2,2k-2l}^{\pi ,0}\left( 0\right)
\int_{0}^{1}dy\left( l+1\right) \,\left( 2l+1\right) \,{\left( \frac{1}{2}%
-\,y\right) }^{2l}\,\,{m(y)}^{2}\log \frac{m(y)^{2}}{\mu ^{2}}\ ,
\label{C00}
\end{equation}%
where the integration variable ${y}$ arises from the Feynman parametrization
and $m(y)=\sqrt{m_{\pi }^{2}-y(1-y)\Delta ^{2}}$. The counterterm
contributions $\gamma _{m+1,2k}^{0}$ and $\zeta _{m+1,2k}^{0}$ depend on $%
\mu $ but not on $m_{\pi }$ and $t$. The $\mu $ dependence of $\gamma
_{m+1,2k}^{0}$ and $\zeta _{m+1,2k}^{0}$ cancels the $\mu $ dependence from $%
C_{M}^{0}$ and $D_{2k+2,2k}^{0,M}$.

As for the $C_{m+1}^{0}$ form factor, it receives non-analytic contributions
from the Fig. 1(d) diagram at $\mathcal{O}(\varepsilon )$ and no
contribution from analytic counterterms until $\mathcal{O}(\varepsilon ^{2})$%
.

\begin{eqnarray}
\mathcal{D}_{2k+2}^{C}\left( t\right) &=&\frac{3g_{A}^{2}M}{32\pi \,F_{\pi
}^{2}}\left\{ \sum_{l=0}^{k}\,A_{2k+2,2k-2l}^{\pi ,0}\left( 0\right)
\int_{0}^{1}dy\,{\left( \frac{1}{2}-\,y\right) }^{2l+2}\left[ \frac{m_{\pi
}^{2}}{m(y)}-4\,\left( l+2\right) \,{m(y)}\right] \right.  \notag \\
&&\left. +\,C_{2k+2}^{\pi ,0}\left( 0\right) \int_{0}^{1}dy\,\left[ \frac{%
m_{\pi }^{2}}{m(y)}-4\,{m(y)}\right] \right\} \ .  \label{C0}
\end{eqnarray}%
By setting $(m,k)=(1,0)$ and using $A_{2,0}^{\pi ,0}\left( 0\right)
=-4C_{2}^{\pi ,0}\left( 0\right) =\left\langle x\right\rangle _{\pi }$, our
results in Eq.(\ref{r1}) reproduce those of Refs. \cite{JqCJ} and \cite{BJ}.

The leading chiral corrections for the isovector ($\alpha =3$) form factors
are

\begin{eqnarray}
E_{m+1,2k}^{3}(t) &=&\left( 1+\varepsilon ^{2}\mathcal{C}_{E}^{3}\right)
E_{m+1,2k}^{3(0)}(0)+\varepsilon ^{2}\delta _{m,2k}\mathcal{D}%
_{2k+1,2k}^{3,E}\left( t\right)  \notag \\
&&+\varepsilon ^{2}\left( 1-\delta _{m,0}\delta _{k,0}\right) \alpha
_{m+1,2k}^{3}m_{\pi }^{2}+\varepsilon ^{2}\beta _{m+1,2k}^{3}t+\mathcal{O}%
(\varepsilon ^{3})\ ,  \notag \\
M_{m+1,2k}^{3}(t) &=&M_{m+1,2k}^{3(0)}(0)+\varepsilon \delta _{m,2k}\mathcal{%
D}_{2k+1,2k}^{3,M}\left( t\right) +\mathcal{O}(\varepsilon ^{2})\ ,  \notag
\\
C_{m+1}^{3}(t) &=&\left( 1+\varepsilon ^{2}\mathcal{C}_{C}^{3}\right)
C_{m+1}^{3(0)}(0)+\varepsilon ^{2}\eta _{m+1}^{3}m_{\pi }^{2}+\varepsilon
^{2}\theta _{m+1}^{3}t+\mathcal{O}(\varepsilon ^{3})\text{ .}  \label{V3}
\end{eqnarray}%
Here%
\begin{eqnarray}
\mathcal{C}_{E}^{3} &=&-\frac{(3g_{A}^{2}+1)m_{\pi }^{2}}{(4\pi F_{\pi })^{2}%
}\ln \frac{m_{\pi }^{2}}{\mu ^{2}},  \notag \\
\mathcal{C}_{C}^{3} &=&-\frac{(3g_{A}^{2}+1)m_{\pi }^{2}}{(4\pi F_{\pi })^{2}%
}\ln \frac{m_{\pi }^{2}}{\mu ^{2}},
\end{eqnarray}%
and%
\begin{eqnarray}
\mathcal{D}_{2k+1,2k}^{3,E}\left( t\right)
&=&\sum_{l=0}^{k}A_{2k+1,2k-2l}^{\pi ,3}\left( 0\right) \frac{\left(
1+2l\right) }{32{\pi }^{2}F_{\pi }^{2}}\,  \notag \\
&&\times \int_{0}^{1}dy{\left( \frac{1}{2}-\,y\right) }^{2l}\left\{
g_{A}^{2}\,\left[ -2\,m_{\pi }^{2}+\left( 5+4l\right) \,{m(y)}^{2}\right] +{%
m(y)}^{2}\right\} \log \frac{m(y)^{2}}{\mu ^{2}}\ ,  \notag \\
\mathcal{D}_{2k+1,2k}^{3,M}\left( t\right)
&=&-\sum_{l=0}^{k}A_{2k+1,2k-2l}^{\pi ,3}\left( 0\right) \frac{Mg_{A}^{2}}{%
8\pi F_{\pi }^{2}}\int_{0}^{1}dy\left\{ \left( 1+2l\right) \,{\left( \frac{1%
}{2}-\,y\right) }^{2l}\,\,m(y)\right\} \ .
\end{eqnarray}%
Here $A_{1,0}^{\pi ,3}\left( 0\right) $ is the number of $u$ quark minus the
number of $d$ quark in a $\pi ^{+}$ meson, $A_{1,0}^{\pi ,3}\left( 0\right)
=\left\langle 1\right\rangle _{u-d}=2$; thus the charge operator $%
E_{1,0}^{3}(0)$ is not renormalized. Unlike the isoscalar case, the chiral
corrections to $M_{m+1,2k}^{3}$ and $C_{m+1}^{3}$ start at $\mathcal{O}%
(\varepsilon )$ and $\mathcal{O}(\varepsilon ^{2})$, respectively.

\section{The Axial Operators}

The Taylor-series expansion of Eq.(\ref{def_2}) gives the form factors of
off-forward nucleon axial twist-2 matrix elements 
\begin{eqnarray}
\langle P^{\prime }|\mathcal{O}_{5,\alpha }^{m}|P\rangle &=&\sum_{\substack{ %
j=0  \\ even}}^{m}\overline{U}(P^{\prime })\left[ n\!\!\!\slash\gamma
_{5}\left( n\cdot \Delta \right) ^{j}\left( n\cdot \bar{P}\right) ^{m-j}%
\tilde{A}_{m+1,j}^{\alpha }(t)\right.  \notag \\
&+&\left. \gamma _{5}\frac{1}{2M}\left( n\cdot \Delta \right) ^{j+1}\left(
n\cdot \bar{P}\right) ^{m-j}\tilde{B}_{m+1,j}^{\alpha }(t)\right] \tau
^{\alpha }U(P)\ ,
\end{eqnarray}%
where%
\begin{equation}
\mathcal{O}_{5,\alpha }^{m}=\bar{q}\tau ^{\alpha }n\!\!\!\slash\gamma
_{5}(n\cdot i\tensor{D})^{m}q\ .  \label{axial}
\end{equation}%
For $m=0$, it reduces to the nucleon axial current matrix element 
\begin{equation}
\langle P^{\prime }|\bar{q}\tau ^{\alpha }\gamma _{\mu }\gamma
_{5}q|P\rangle =\overline{U}(P^{\prime })\left[ \gamma _{\mu }G_{A}(t)+\frac{%
\Delta _{\mu }}{2M}G_{P}(t)\right] \gamma _{5}\tau ^{a}U(P)\ ,
\end{equation}%
with $\tilde{A}_{1,0}=G_{A}$ and $\tilde{B}_{1,0}=G_{P}$. The form factors
are related to the moments of axial GPDs as 
\begin{eqnarray}
\tilde{H}_{m+1} &=&\int_{-1}^{1}dxx^{m}\tilde{H}(x,\xi ,t)=\sum_{\substack{ %
j=0  \\ even}}^{m}(-2\xi )^{j}~\tilde{A}_{m+1,j}(t)\ ,  \notag \\
\tilde{E}_{m+1} &=&\int_{-1}^{1}dxx^{m}\tilde{E}(x,\xi ,t)=\sum_{\substack{ %
j=0  \\ even}}^{m}(-2\xi )^{j}~\tilde{B}_{m+1,j}(t)\ .  \label{aaa}
\end{eqnarray}

Similar to the vector twist-2 operators, $O_{5,\alpha }^{m}$ is matched to
pionic and nucleon operators%
\begin{equation}
\mathcal{O}_{5,\alpha }^{m}\,\rightarrow O_{5,\alpha ,\pi }^{m}+O_{5,\alpha
,N}^{m}+\cdots \ .
\end{equation}%
Since $O_{5,\alpha }^{m}=O_{\alpha ,R}^{m}-O_{\alpha ,L}^{m}$, the matching
to leading pionic operators is similar to that of Eq.(\ref{pi-mat}) with a
different relative sign,

\begin{eqnarray}
O_{5,\alpha ,\pi }^{m} &=&\frac{F_{\pi }^{2}}{4}\sum_{\substack{ j=0  \\ %
even }}^{m}\overline{A}_{m+1,j}^{\pi ,\alpha }(0)\left( -in\cdot \partial
\right) ^{j}tr\left[ -\tau _{L}^{\alpha }\!\!\ \Sigma \left( in\cdot %
\tensor{\partial}\right) ^{m-j+1}\Sigma ^{\dagger }\right.  \notag \\
&&\left. +\tau _{R}^{\alpha }\!\!\ \Sigma ^{\dagger }\left( in\cdot %
\tensor{\partial}\right) ^{m-j+1}\Sigma \right] +\cdots \   \label{O_5}
\end{eqnarray}%
Parity conservation governs that the axial operators match onto pionic
operators with odd number of pions; thus diagrams in Figs. 1(d) and (e)
could not contribute. Instead, the pion pole diagram in Fig. 2 gives the
leading contribution from insertion of $O_{5,\alpha ,\pi }^{m}$. In this
diagram, a necessary input is the $\pi \rightarrow 0$ matrix element of $%
O_{5,\alpha ,\pi }^{m}$ which is related to the $(m+1)$-th moment of pion
light cone distribution function 
\begin{equation}
\langle \pi ^{b}(\Delta )|\mathcal{O}_{5,\alpha }^{m}|0\rangle =i\delta
_{ab}F_{\pi }\left( -n\cdot \Delta \right) ^{m+1}\left\langle
z^{m}\right\rangle _{\pi }/2^{m}\ .  \label{pi->0}
\end{equation}%
where $z=1-2x^{\prime }$, 
\begin{equation}
\left\langle z^{m}\right\rangle _{\pi }=\int_{0}^{1}dx^{\prime }\left(
1-2x^{\prime }\right) ^{m}\phi _{\pi }\left( x^{\prime }\right) \ ,
\label{LCWF}
\end{equation}%
and where $\phi _{\pi }\left( x^{\prime }\right) $ is the pion light cone
distribution function with the normalization $\left\langle
z^{0}\right\rangle _{\pi }=\int_{0}^{1}dx^{\prime }\phi _{\pi }\left(
x^{\prime }\right) =1$. $\left\langle z^{m}\right\rangle _{\pi }$ vanishes
for odd $m$ due to charge conjugation. In Ref. \cite{CS}, it was shown that
the leading non-analytic quark mass corrections to $\left\langle
z^{m}\right\rangle $ can all be absorbed into $F_{\pi }$. Thus $\left\langle
z^{m}\right\rangle _{\pi }$ is purely analytic at $\mathcal{O}(\varepsilon
^{2})$ \cite{CS} but not analytic at higher orders \cite{t3}. Using the
operators constructed in Eqs. (\ref{pi-mat}) and (\ref{O_5}), our
leading-order (LO) result reproduces that of Ref. \cite{Polyakov:1998ze}
derived from the soft pion theorem, 
\begin{equation}
H_{\pi }^{\alpha =3}(x,\xi _{\pi }=1,t=0)=\frac{1}{2}\phi _{\pi }\left( 
\frac{1+x}{2}\right) \ .
\end{equation}

In heavy baryon $\chi $PT, the leading axial nucleon operators in the
matching are 
\begin{eqnarray}
O_{5,\alpha ,N}^{m} &=&\sum_{\substack{ j=0  \\ even}}^{m}2M\overline{N}%
\left[ 2n\cdot S\left( n\cdot \Delta \right) ^{j}\left( Mn\cdot v\right)
^{m-j}\tilde{E}_{m+1,j}^{\alpha (0)}(0)\right.  \notag \\
&+&\left. \frac{(S\cdot \Delta )}{2M^{2}}\left( n\cdot \Delta \right)
^{j+1}\left( Mn\cdot v\right) ^{m-j}\tilde{M}_{m+1,j}^{\alpha (0)}(0)\right]
\tau _{u+}^{\alpha }N+\cdots \ ,  \label{HBaxial}
\end{eqnarray}%
\newline
where%
\begin{equation}
\tilde{E}_{m+1,j}=\tilde{A}_{m+1,j}\ ,\quad \tilde{M}_{m+1,j}=\frac{\tilde{A}%
_{m+1,j}}{1+\sqrt{1-\frac{t}{4M^{2}}}}+\tilde{B}_{m+1,j}\ .
\end{equation}

\begin{figure}[t]
\vskip 0.cm 
\centerline{
  \mbox{\epsfxsize=5.0truecm \hbox{\epsfbox{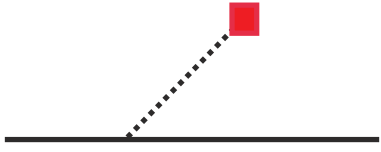}} }
  } \vskip-0.1cm \vspace{-0.5cm} 
\caption[1]{The leading order pion pole diagram contributing to the nucleon
pseudo-scalar form factors. }
\label{Fig2}
\end{figure}

For the isoscalar case, the leading chiral corrections include diagrams of
Figs. 1(a), (b), (f):%
\begin{eqnarray}
\tilde{E}_{m+1,2k}^{0}(t) &=&\left( 1+\varepsilon ^{2}\mathcal{C}%
_{M}^{0}\right) \tilde{E}_{m+1,2k}^{0(0)}(0)+\varepsilon ^{2}\widetilde{%
\alpha }_{m+1,2k}^{0}m_{\pi }^{2}+\varepsilon ^{2}\widetilde{\beta }%
_{m+1,2k}^{0}t+\mathcal{O}(\varepsilon ^{3})\ ,  \label{a1} \\
\tilde{M}_{m+1,2k}^{0}(t) &=&\left( 1+\varepsilon ^{2}\mathcal{C}%
_{M}^{0}\right) \tilde{M}_{m+1,2k}^{0(0)}(0)+\varepsilon ^{2}\frac{\mathcal{C%
}_{M}^{0}}{3}\tilde{E}_{m+1,2k}^{0(0)}(0)+\varepsilon ^{2}\widetilde{\gamma }%
_{m+1,2k}^{0}m_{\pi }^{2}+\varepsilon ^{2}\widetilde{\zeta }_{m+1,2k}^{0}t+%
\mathcal{O}(\varepsilon ^{3})\ .  \notag
\end{eqnarray}%
The contribution of $\tilde{E}_{m+1,2k}^{0(0)}$ to $\tilde{M}_{m+1,2k}^{0}$
is from $1/M^{2\text{ }}$corrections of Fig. 1(b) type diagrams \cite%
{Diehl:2006ya}.

For the isovector case, the leading non-analytic contribution of $\tilde{E}%
^{3}$ is from Figs. 1(b), (c) and (f).%
\begin{equation}
\tilde{E}_{m+1,2k}^{3}(t)=\left( 1+\varepsilon ^{2}\mathcal{C}%
_{M}^{3}\right) \tilde{E}_{m+1,2k}^{3(0)}(0)+\varepsilon ^{2}\widetilde{%
\alpha }_{m+1,2k}^{3}m_{\pi }^{2}+\varepsilon ^{2}\widetilde{\beta }%
_{m+1,2k}^{3}t+\mathcal{O}(\varepsilon ^{3})\ ,  \label{a2}
\end{equation}%
where%
\begin{equation}
\mathcal{C}_{M}^{3}=-\frac{(2g_{A}^{2}+1)m_{\pi }^{2}}{(4\pi F_{\pi })^{2}}%
\ln \frac{m_{\pi }^{2}}{\mu ^{2}}\ .
\end{equation}%
The $\mu $ dependence in Eqs. (\ref{a1}) and (\ref{a2}) is absorbed by the
counterterms $\widetilde{\alpha }_{m+1,2k}^{0}$, $\widetilde{\gamma }%
_{m+1,2k}^{0}$ and $\widetilde{\alpha }_{m+1,2k}^{3}$. \ 

In contrary, the leading contribution to $\tilde{M}^{3}$ is non-analytic [$%
\mathcal{O}(\varepsilon ^{-2})$] arising from the pion pole diagram shown in
Fig. 2. At one loop, we obtain 
\begin{equation}
\tilde{M}_{m+1,2k}^{3}(t)=\delta _{m,2k}G_{P}(t)\left\langle
z^{2k}\right\rangle _{\pi }/2^{2k}+\tilde{M}_{m+1,2k}^{3(0)}(0)+\mathcal{O}%
(\varepsilon )\ ,  \label{tildeB}
\end{equation}%
where $\left\langle z^{2k}\right\rangle _{\pi }$ is analytic at $\mathcal{O}%
(\varepsilon ^{2})$ as mentioned above and the pseudo-scalar form factor 
\cite{Bernard:2001rs}, 
\begin{equation}
G_{P}(t)=\frac{4g_{A}M^{2}}{m_{\pi }^{2}-t}\left( \frac{1}{\varepsilon ^{2}}-%
\frac{2\overline{d}_{18}}{g_{A}}m_{\pi }^{2}\right) -\frac{2}{3}%
g_{A}M^{2}\left\langle r_{A}^{2}\right\rangle \ ,  \label{Gp}
\end{equation}%
where $\overline{d}_{18}$ is a counterterm and $\left\langle
r_{A}^{2}\right\rangle $ is the square of nucleon axial charge radius.

\section{The Tensor Operator}

The tensor operator

\begin{equation}
\mathcal{O}_{T,\alpha }^{m}=\bar{q}\tau ^{\alpha }in_{\mu }r_{\nu }\sigma
^{\mu \nu }(in\cdot D)^{m}q
\end{equation}%
has the nucleon matrix element \cite{Hagler:2004yt}%
\begin{eqnarray}
\langle P^{\prime }|\mathcal{O}_{T,\alpha }^{m}|P\rangle &=&\sum_{j=0}^{m}%
\overline{U}(P^{\prime })\left( n\cdot \Delta \right) ^{j}\left( n\cdot \bar{%
P}\right) ^{m-j}\left[ in_{\mu }r_{\nu }\sigma ^{\mu \nu }\left. {A}%
_{m+1,j}^{T,\alpha }(t)\right\vert _{j\ even}\right.  \notag \\
&+&\frac{\left( n\cdot \bar{P}\right) \left( r\cdot \Delta \right) -\left(
r\cdot \bar{P}\right) \left( n\cdot \Delta \right) }{M^{2}}\left. \tilde{A}%
_{m+1,j}^{T,\alpha }(t)\right\vert _{j\ even}  \notag \\
&+&\frac{n\!\!\!\slash\left( r\cdot \Delta \right) -r\!\!\!\slash\left(
n\cdot \Delta \right) }{2M}\left. {B}_{m+1,j}^{T,\alpha }(t)\right\vert _{j\
even}  \notag \\
&+&\left. \frac{n\!\!\!\slash\left( r\cdot \bar{P}\right) -r\!\!\!\slash%
\left( n\cdot \bar{P}\right) }{M}\left. \tilde{B}_{m+1,j}^{T,A}(t)\right%
\vert _{j\ odd}\right] \tau ^{\alpha }U(P)\ .  \label{tensor}
\end{eqnarray}%
The form factors are related to the moments of nucleon tensor GPDs as 
\begin{eqnarray}
{H}_{T,m+1} &=&\int_{-1}^{1}dxx^{m}H_{T}(x,\xi ,t)=\sum_{j=0,even}^{m}(-2\xi
)^{j}~A_{m+1,j}^{T}(t)\ ,  \notag \\
E_{T,m+1} &=&\int_{-1}^{1}dxx^{m}E_{T}(x,\xi ,t)=\sum_{j=0,even}^{m}(-2\xi
)^{j}~B_{m+1,j}^{T}(t)\ ,  \notag \\
\tilde{H}_{T,m+1} &=&\int_{-1}^{1}dxx^{m}\tilde{H}_{T}(x,\xi
,t)=\sum_{j=0,even}^{m}(-2\xi )^{j}~\tilde{A}_{m+1,j}^{T}(t)\ ,  \notag \\
\tilde{E}_{T,m+1} &=&\int_{-1}^{1}dxx^{m}\tilde{E}_{T}(x,\xi
,t)=\sum_{j=0,odd}^{m}(-2\xi )^{j}~\tilde{B}_{m+1,j}^{T}(t)\ .  \label{aaaa}
\end{eqnarray}%
After the $1/M$ expansions, Eq. (\ref{tensor}) becomes 
\begin{eqnarray}
\langle P^{\prime }|\mathcal{O}_{T,\alpha }^{m}|P\rangle &=&\sum_{j=0}^{m}2M%
\overline{N}(P^{\prime })\left( n\cdot \Delta \right) ^{j}\left( Mn\cdot
v\right) ^{m-j}\left[ 2i{\epsilon }^{vnrS}\left. {M}_{m+1,j}^{T,\alpha
}(t)\right\vert _{j\ even}\right.  \notag \\
&+&\frac{\left( n\cdot v\right) \left( r\cdot \Delta \right) }{M}\left.
E_{m+1,j}^{T,\alpha }(t)\right\vert _{j\ even}-\frac{i\left( n\cdot v\right) 
{\epsilon }^{vr\Delta S}}{M}\left. {C}_{m+1,j}^{T,\alpha }(t)\right\vert
_{j\ odd}  \notag \\
&+&\left. i\frac{\left( n\cdot \Delta \right) {\epsilon }^{vr\Delta
S}-\left( r\cdot \Delta \right) {\epsilon }^{vn\Delta S}}{8M^{2}}\left.
W_{m+1,j}^{T,A}(t)\right\vert _{j\ even}\right] \tau ^{\alpha }N(P)\ ,
\end{eqnarray}%
with 
\begin{eqnarray}
M_{m+1,j}^{T} &=&\left( 1-\frac{t}{16M^{2}}\right) A_{m+1,j}^{T}\
,\,\,\,E_{m+1,j}^{T}=\frac{1}{2}A_{m+1,j}^{T}+\tilde{A}_{m+1,j}^{T}+\frac{1}{%
2}B_{m+1,j}^{T}\ ,  \notag \\
C_{m+1,j}^{T} &=&\tilde{B}_{m+1,j}^{T}\
,\,\,\,\,W_{m+1,j}^{T}=-2A_{m+1,j}^{T}-4B_{m+1,j}^{T}\ .  \label{73}
\end{eqnarray}%
Under charge conjugation,%
\begin{equation}
\mathcal{CO}_{T,\alpha }^{m}\mathcal{C}^{-1}=\left( -1\right) ^{m+1}\mathcal{%
O}_{T,\alpha }^{m}\ \quad \left( \alpha =0,3\right) \ .
\end{equation}%
By similar arguments to the vector operators, the moments of pionic tensor
GPD can be parametrized as%
\begin{eqnarray}
&&\langle \pi ^{i}(P_{\pi }^{\prime })|\bar{q}(in_{\mu }r_{\nu }\sigma ^{\mu
\nu })(in\cdot D)^{2k+1}q|\pi ^{j}(P_{\pi })\rangle  \notag \\
&=&2\delta _{ij}\sum_{l=0}^{k}\frac{1}{M}\left[ \left( n\cdot \overline{P}%
_{\pi }\right) \left( r\cdot \Delta \right) -\left( n\cdot \Delta \right)
\left( r\cdot \overline{P}_{\pi }\right) \right] \left( n\cdot \Delta
\right) ^{2l}\left( n\cdot \overline{P}_{\pi }\right)
^{2k-2l+1}E_{2k+2,2l}^{T,\pi ,0}(t) \\
&&\langle \pi ^{i}(P_{\pi }^{\prime })|\bar{q}\tau ^{3}(in_{\mu }r_{\nu
}\sigma ^{\mu \nu })(in\cdot D)^{2k}q|\pi ^{j}(P_{\pi })\rangle  \notag \\
&=&2i\epsilon ^{i3j}\sum_{l=1}^{k}\frac{1}{M}\left[ \left( n\cdot \overline{P%
}_{\pi }\right) \left( r\cdot \Delta \right) -\left( n\cdot \Delta \right)
\left( r\cdot \overline{P}_{\pi }\right) \right] \left( n\cdot \Delta
\right) ^{2l}\left( n\cdot \overline{P}_{\pi }\right)
^{2k-2l}E_{2k+1,2l}^{T,\pi ,3}(t)
\end{eqnarray}%
where we have inserted a heavy scale $M$ which is of order $\mu _{\chi }$ to
make $E_{m,j}^{T,\pi }$ dimensionless.

The matching procedure is similar to the vector and axial vector cases which
we will not repeat here. We just point out one main difference. $O_{T,\alpha
}^{m}$ can be decomposed as 
\begin{equation}
\mathcal{O}_{T,\alpha }^{m}=\bar{q}_{L}\tau _{LR}^{\alpha }in_{\mu }r_{\nu
}\sigma ^{\mu \nu }(in\cdot D)^{m}q_{R}+\bar{q}_{R}\tau _{RL}^{\alpha
}in_{\mu }r_{\nu }\sigma ^{\mu \nu }(in\cdot D)^{m}q_{L}\ .
\end{equation}%
We will set $\tau _{LR}^{\alpha }=\tau _{RL}^{\alpha }=\tau ^{\alpha }$ at
the end. $O_{T,\alpha }^{m}$ will be invariant under a chiral rotation if we
demand 
\begin{equation}
\tau _{LR}^{\alpha }\rightarrow L\tau _{LR}^{\alpha }R^{\dagger }\ ,\quad
\tau _{RL}^{\alpha }\rightarrow R\tau _{RL}^{\alpha }L^{\dagger }\ .
\end{equation}%
Thus instead of using $\tau _{u\pm }^{a}$, we use $\tau _{\pm }^{a}=\frac{1}{%
2}\left( u^{\dagger }\tau _{LR}^{a}u^{\dagger }\pm u\tau _{RL}^{a}u\right) $
which transforms as $\tau _{\pm }^{a}\rightarrow U(x)\tau _{\pm
}^{a}U(x)^{\dagger }$ under a chiral rotation to construct the nucleon
operators \cite{CJ}.

For the isoscalar form factors,%
\begin{eqnarray}
{M}_{m+1,2k}^{T,0}(t) &=&\left( 1+\varepsilon ^{2}\mathcal{C}_{M}^{0}\right) 
{M}_{m+1,2k}^{T,0(0)}(0)+\varepsilon ^{2}\widetilde{\alpha }%
_{m+1,2k}^{T,0}m_{\pi }^{2}+\varepsilon ^{2}\widetilde{\beta }%
_{m+1,2k}^{T,0}t+\mathcal{O}(\varepsilon ^{3})\ ,  \notag \\
{E}_{m+1,2k}^{T,0}(t) &=&{E}_{m+1,2k}^{T,0(0)}(0)+\varepsilon ^{2}\widetilde{%
\gamma }_{m+1,2k}^{T,0}m_{\pi }^{2}+\varepsilon ^{2}\widetilde{\zeta }%
_{m+1,2k}^{T,0}t+\mathcal{O}(\varepsilon ^{3})\ ,  \notag \\
{C}_{m+1,2k+1}^{T,0}(t) &=&\left( 1+\varepsilon ^{2}\mathcal{C}%
_{M}^{0}\right) {C}_{m+1,2k+1}^{T,0(0)}(0)+\varepsilon ^{2}\delta _{m,2k+1}%
\mathcal{T}_{2k+2,2k+1}^{0,C}(t)  \notag \\
&&+\varepsilon ^{2}\widetilde{\eta }_{m+1,2k}^{T,0}m_{\pi }^{2}+\varepsilon
^{2}\widetilde{\theta }_{m+1,2k}^{T,0}t+\mathcal{O}(\varepsilon ^{3})\ , 
\notag \\
{W}_{m+1,2k}^{T,0}(t) &=&\left( 1+\varepsilon ^{2}\mathcal{C}_{M}^{0}\right) 
{W}_{m+1,2k}^{T,0(0)}(0)+\varepsilon ^{2}\frac{4}{3}\mathcal{C}_{M}^{0}{M}%
_{m+1,2k}^{T,0(0)}(0)+\varepsilon ^{2}\delta _{m,2k+1}\mathcal{T}%
_{2k+2,2k}^{0,W}(t)  \notag \\
&&+\varepsilon ^{2}\widetilde{\kappa }_{m+1,2k}^{T,0}m_{\pi
}^{2}+\varepsilon ^{2}\widetilde{\lambda }_{m+1,2k}^{T,0}t+\mathcal{O}%
(\varepsilon ^{3})\ ,
\end{eqnarray}%
where%
\begin{eqnarray}
\mathcal{T}_{2k+2,2k+1}^{0,C}(t) &=&-\frac{3g_{A}^{2}}{32{\pi }^{2}F_{\pi
}^{2}}\sum_{l=0}^{k}E_{2k+2,2k-2l}^{T,\pi ,0}(0)\left( 2l+1\right)
\int_{0}^{1}dy\,{\left( \frac{1}{2}-\,y\right) }^{2l}{m(y)}^{2}\log \frac{%
m(y)^{2}}{\mu ^{2}}\ ,  \notag \\
\mathcal{T}_{2k+2,2k}^{0,W}(t) &=&16\mathcal{T}_{2k+2,2k+1}^{0,C}(t)\ .
\end{eqnarray}%
Similar to Eq. (\ref{a1}), the contribution of ${M}_{m+1,2k}^{T,0(0)}$ to ${W%
}_{m+1,2k}^{T,0(0)}$ is from $1/M^{2\text{ }}$corrections of Fig. 1(b) type
diagrams.

Similarly, for the isovector form factors,

\begin{eqnarray}
{M}_{m+1,2k}^{T,3}(t) &=&\left( 1+\varepsilon ^{2}\mathcal{C}%
_{M}^{3,T}\right) {M}_{m+1,2k}^{T,3(0)}(0)+\varepsilon ^{2}\widetilde{\alpha 
}_{m+1,2k}^{T,3}m_{\pi }^{2}+\varepsilon ^{2}\widetilde{\beta }%
_{m+1,2k}^{T,3}t+\mathcal{O}(\varepsilon ^{3})\ ,  \notag \\
{E}_{m+1,2k}^{T,3}(t) &=&\left( 1+\varepsilon ^{2}\mathcal{C}_{E}^{3}\right) 
{E}_{m+1,2k}^{T,3(0)}(0)+\varepsilon ^{2}\delta _{m,2k}\mathcal{T}%
_{2k+1,2k}^{3,E}(t)+\varepsilon ^{2}\widetilde{\gamma }_{m+1,2k}^{T,3}m_{\pi
}^{2}+\varepsilon ^{2}\widetilde{\zeta }_{m+1,2k}^{T,3}t+\mathcal{O}%
(\varepsilon ^{3})\ ,  \notag \\
{C}_{m+1,2k+1}^{T,3}(t) &=&\left( 1+\varepsilon ^{2}\mathcal{C}%
_{M}^{3,T}\right) {C}_{m+1,2k+1}^{T,3(0)}(0)+\varepsilon ^{2}\widetilde{\eta 
}_{m+1,2k}^{T,3}m_{\pi }^{2}+\varepsilon ^{2}\widetilde{\theta }%
_{m+1,2k}^{T,3}t+\mathcal{O}(\varepsilon ^{3})\ ,  \notag \\
{W}_{m+1,2k}^{T,3}(t) &=&{W}_{m+1,2k}^{T,3(0)}(0)+\varepsilon \delta _{m,2k}%
\mathcal{T}_{2k+1,2k}^{3,W}(t)+\mathcal{O}(\varepsilon ^{2})\ ,  \label{79}
\end{eqnarray}%
where%
\begin{equation}
\mathcal{C}_{M}^{3,T}=-\frac{2g_{A}^{2}+1/2}{(4\pi F_{\pi })^{2}}m_{\pi
}^{2}\ln \frac{m_{\pi }^{2}}{\mu ^{2}}\ ,
\end{equation}%
and%
\begin{eqnarray}
\mathcal{T}_{2k+1,2k}^{3,E}(t) &=&-\frac{1}{32{\pi }^{2}F_{\pi }^{2}}%
\sum_{l=0}^{k}E_{2k+1,2k-2l}^{T,\pi ,3}(0)\int_{0}^{1}dy\,{\left( \frac{1}{2}%
-\,y\right) }^{2l}  \notag \\
&&\times \left\{ g_{A}^{2}\left[ 2m^{2}-\left( 5+4l\right) {m(y)}^{2}\right]
-{m(y)}^{2}\right\} \log \frac{m(y)^{2}}{\mu ^{2}}  \notag \\
\mathcal{T}_{2k+1,2k}^{3,W}(t) &=&\frac{Mg_{A}^{2}}{{\pi }F_{\pi }^{2}}%
\sum_{l=0}^{k}E_{2k+1,2k-2l}^{T,\pi ,3}(0)\int_{0}^{1}dy{\left( \frac{1}{2}%
-\,y\right) }^{2l}{m(y)}\,\,
\end{eqnarray}

It is interesting that the leading chiral corrections of the tensor and
vector matrix elements are similar. They obey the relations,%
\begin{eqnarray}
\mathcal{T}_{2k+2,2k+1}^{0,C}(t) &=&-\frac{1}{2}\left. \mathcal{D}%
_{2k+2,2k}^{0,M}\left( t\right) \right\vert _{\,\left( l+1\right)
A_{2k+2,2k-2l}^{\pi ,0}\rightarrow E_{2k+2,2k-2l}^{T,\pi ,0}}\ ,  \notag \\
\mathcal{T}_{2k+1,2k}^{3,E}(t) &=&\left. \mathcal{D}_{2k+1,2k}^{3,E}\left(
t\right) \right\vert _{\,\left( 2l+1\right) A_{2k+1,2k-2l}^{\pi
,3}\rightarrow E_{2k+1,2k-2l}^{T,\pi ,3}}\ ,  \notag \\
\mathcal{T}_{2k+1,2k}^{3,W}(t) &=&8\left. \mathcal{D}_{2k+1,2k}^{3,M}\left(
t\right) \right\vert _{\,\left( 2l+1\right) A_{2k+1,2k-2l}^{\pi
,3}\rightarrow E_{2k+1,2k-2l}^{T,\pi ,3}}\ .
\end{eqnarray}

\section{Gluon GPDs}

The gluon GPDs are defined as \cite{Hoodbhoy:1998vm,Diehl:2001pm}

\begin{eqnarray*}
n_{\mu }n_{\nu }\langle P^{\prime }|\,F^{\mu \alpha }\left( -\frac{z}{2}%
n\right) \!\!F_{\alpha }^{\ \nu }\left( \frac{z}{2}n\right) \,\,|P\rangle
&=&\int dxe^{-ixzn\cdot \bar{P}}\overline{U}(P^{\prime })n\cdot \overline{P}%
\left[ xH_{g}n\!\!\!\slash+xE_{g}\,\frac{i\sigma ^{\alpha \beta }n_{\alpha
}\Delta _{\beta }}{2M}\right] U(P)\ , \\
n_{\mu }n_{\nu }\langle P^{\prime }|\,F^{\mu \alpha }\left( -\frac{z}{2}%
n\right) i\!\!\ \tilde{F}_{\alpha }^{\ \nu }\left( \frac{z}{2}n\right)
\,\,|P\rangle &=&\int dxe^{-ixzn\cdot \bar{P}}\overline{U}(P^{\prime
})n\cdot \overline{P}\left[ x\tilde{H}_{g}n\!\!\!\slash\gamma _{5}+x\tilde{E}%
_{g}\,\frac{\gamma _{5}n\cdot \Delta }{2M}\right] U(P)\ ,
\end{eqnarray*}%
\begin{eqnarray}
\lefteqn{n_{\mu }n_{\nu }r_{\alpha }r_{\beta }\langle P^{\prime }|\,F^{\mu
\alpha }\left( -\frac{z}{2}n\right) \!\!F^{\nu \beta }\left( \frac{z}{2}%
n\right) \,\,|P\rangle }  \notag \\
&=&\int dxe^{-ixzn\cdot \bar{P}}\left[ \left( n\cdot \overline{P}\right)
\left( r\cdot \Delta \right) -\left( n\cdot \Delta \right) \left( r\cdot 
\overline{P}\right) \right]  \notag \\
&&\times \overline{U}(P^{\prime })\left[ xH_{Tg}\,in_{\alpha }r_{\beta
}\sigma ^{\alpha \beta }+x\tilde{H}_{Tg}\,\frac{\left( n\cdot \overline{P}%
\right) \left( r\cdot \Delta \right) -\left( r\cdot \overline{P}\right)
\left( n\cdot \Delta \right) \,}{M^{2}}\right.  \notag \\
&&\left. +xE_{Tg}\,\frac{n\!\!\!\slash\left( r\cdot \Delta \right) -r\!\!\!%
\slash\left( n\cdot \Delta \right) \,}{2M}+x\tilde{E}_{Tg}\,\frac{n\!\!\!%
\slash\left( r\cdot \overline{P}\right) -r\!\!\!\slash\left( n\cdot 
\overline{P}\right) \,}{2M}\right] U(P)\ ,  \label{H_g}
\end{eqnarray}%
where the Wilson lines are also suppressed. The Taylor-series expansion in $%
z $ of the above equations gives rise to relations between the gluon twist-2
matrix elements and the moments of gluon GPDs: 
\begin{equation}
F_{g,m+1}(\xi ,t)=\int dxx^{m}F_{g}(x,\xi ,t)\ ,
\end{equation}%
where $F_{g}(x,\xi ,t)$ denotes a generic gluon GPD. Here, unlike the quark
case, $m=1,2,\ldots $without $m=0$. This is because the right-hand sides of
Eq. (\ref{H_g}) are of the form $xF_{g}$ instead of $F_{g}$. Therefore $%
F_{g} $ can only be determined up to a function of the form $\lambda (\xi
,t)\delta (x)$ using the above definitions. However, the first moments in $x$
of gluon GPDs can be probed by non-local gauge invariant operators as
introduced in Ref. \cite{Manohar}.

The local gluon twist-2 operators and isoscalar quark twist-2 operators all
transform in the same way---as singlets---under chiral transformation. The
gluon twist-2 operators match onto the same set of hadronic operators as the
isoscalar quark twist-2 operators with different prefactors. Thus our
previous results for the moments of isoscalar quark GPDs can be easily
converted to moments of gluon GPDs.

\section{$\protect\chi $PT constraints on the Nucleon GPDs}

In this section we use the moments of nucleon GPDs calculated above to shed
light on the GPDs themselves. We first pay attention to the non-analytic $%
\mathcal{O}(\varepsilon )$ corrections before making transit to the impact
parameter distributions of GPDs. There are no counterterms at $\mathcal{O}%
(\varepsilon )$, therefore those corrections are clean predictions of $\chi $%
PT. Here are several interesting cases of this type:

1) At $\mathcal{O}(\varepsilon )$, $E^{0}$ and $H^{0}$ only receive chiral
corrections from the \textquotedblleft D terms\textquotedblright\ which only
depend on $x/\xi $ and $t$ \cite{Polyakov:1999gs}, 
\begin{equation}
\delta E^{0}(x,\xi ,t)=-\delta H^{0}(x,\xi ,t)=D(\frac{x}{\xi },t)\theta
\left( 1-\left\vert \frac{x}{\xi }\right\vert \right) \ ,  \label{11}
\end{equation}%
with $D(z,t)=-D(-z,t)$. From Eqs.(\ref{aa}), (\ref{r1}) and (\ref{C0}),

\begin{equation}
\delta H_{m+1}^{0}(\xi ,t)=\int_{-1}^{+1}dxx^{m}\delta H^{0}(x,\xi
,t)=(-2\xi )^{m+1}C_{m+1}(t)\ ,  \label{12}
\end{equation}%
where $C_{m+1}(t)=0$ for $m$ even. Then one can prove $\delta E^{0}$ and $%
\delta H^{0}$ have the functional form in Eq.(\ref{11}).

2) $E^{3}$, $E_{T}^{3}$ and $\tilde{H}_{T}^{3}$ receive $\mathcal{O}%
(\varepsilon )$ contributions which are predictable. We will discuss their
impact parameter distributions in the next subsection.

3) One can show that Eq.(\ref{tildeB}) implies 
\begin{equation}
\tilde{E}^{3}(x,\xi ,t)=\frac{\phi _{\pi }\left( \frac{\xi -x}{2\xi }\right) 
}{2\left\vert \xi \right\vert }G_{P}(t)+\mathcal{O}(\varepsilon ^{0})\ .
\label{tildeE3}
\end{equation}%
This result coincides with those derived in Refs. \cite%
{Mankiewicz:1998kg,Penttinen:1999th} after using $\phi _{\pi }(x^{\prime
})=\phi _{\pi }(1-x^{\prime })$.

\subsection{Impact Parameter Distributions}

\label{IPD}

In the limit of $\xi \rightarrow 0$ (and $t\rightarrow -\Delta _{\perp }^{2}$%
), the $\Delta _{\perp }$ Fourier transformation of the GPDs has the
interpretation of simultaneous measurement of the longitudinal momentum and
transverse position (impact parameter) of partons in the infinite momentum
frame. The impact parameter dependent parton distribution for a generic
nucleon GPD $F$ is 
\begin{equation}
\mathcal{F}(x,b_{\perp })=\int \frac{d^{2}\mathbf{\Delta }_{\perp }}{\left(
2\pi \right) ^{2}}e^{i\mathbf{b}_{\perp }\cdot \mathbf{\Delta }_{\perp
}}F(x,0,-\mathbf{\Delta }_{\perp }^{2})\ .
\end{equation}%
We are interested in the small $t$ ($t\ll \mu _{\chi }^{2}$), or large $%
b_{\perp }$($b\gg 1/\mu _{\chi }$) region where $\chi $PT is reliable, so we
expand $F$ in $t$,

\begin{equation}
F(x,0,t=-\mathbf{\Delta }_{\perp }^{2})=F(x,0,0)-\partial _{t}F(x,0,0)%
\mathbf{\Delta }_{\perp }^{2}+\cdots \ ,
\end{equation}%
which leads to 
\begin{equation}
\mathcal{F}(x,b_{\perp })=F(x,0,0)\delta ^{2}\left( \mathbf{b}_{\perp
}\right) +\partial _{t}F(x,0,0)\nabla _{\perp }^{2}\delta ^{2}\left( \mathbf{%
b}_{\perp }\right) +\cdots \ .
\end{equation}%
The $F$ term gives the $b_{\perp }$ integrated distribution 
\begin{equation}
\int d^{2}\mathbf{b}_{\perp }\mathcal{F}(x,b_{\perp })=F(x,0,0)\ ,
\end{equation}%
while the $\partial _{t}F$ term is related to the averaged $b_{\perp }^{2}$
of $F$ as a function of $x$, 
\begin{equation}
\left\langle b_{\perp }^{2}\right\rangle _{\mathcal{F}}=\frac{\int d^{2}%
\mathbf{b}_{\perp }b_{\perp }^{2}\mathcal{F}(x,b_{\perp })}{\int d^{2}%
\mathbf{b}_{\perp }\mathcal{F}(x,b_{\perp })}=4\frac{\partial _{t}F(x,0,0)}{%
F(x,0,0)}\ .  \label{b_per}
\end{equation}%
The terms with higher derivatives on the delta function give higher moments
of $b_{\perp }^{2}$. In comparison, the charge radii of the electroweak form
factors defined in Eq. (\ref{F1F2}) are%
\begin{equation}
\left\langle r^{2}\right\rangle _{F}=6\frac{\int dx\partial _{t}F(x,0,0)}{%
\int dxF(x,0,0)}\ .  \label{R}
\end{equation}%
They also constrain the functional form of $F$.

In the following paragraphs, we will extract from the model-independent
results of twist-2 matrix elements we obtained above to obtain $F(x,0,0)$
and $\left\langle b_{\perp }^{2}\right\rangle _{\mathcal{F}}$. The GPD $%
\tilde{E}^{3}(x,\xi ,t)$ has a special $\xi \rightarrow 0$ limit:%
\begin{eqnarray}
\tilde{E}^{3}(x,0,t) &=&\delta (x)G_{P}(t)+\mathcal{O}(\varepsilon ^{0}) 
\notag \\
&=&\delta (x)\frac{4g_{A}M^{2}}{m_{\pi }^{2}-t}\frac{1}{\varepsilon ^{2}}+%
\mathcal{O}(\varepsilon ^{0})\ ,
\end{eqnarray}%
where we have used Eqs.(\ref{tildeE3}) and (\ref{Gp}), $\phi _{\pi }\left(
x^{\prime }\right) =0$ for $x^{\prime }<0$ or $x^{\prime }>1$, and $%
\int_{0}^{1}dx^{\prime }\phi _{\pi }\left( x^{\prime }\right) =1$. This
yields $\left\langle b_{\perp }^{2}\right\rangle _{\mathcal{\tilde{E}}%
^{3}}=4/m_{\pi }^{2}$ at $x=0$.

For the other GPDs, it is convenient to express $F(x,0,0)$ and $\partial
_{t}F(x,0,0)$ as (except $\tilde{E}^{3}$ which we will discuss later) 
\begin{eqnarray}
F(x,0,0) &=&a_{F}(x)+\varepsilon b_{F}(x)\frac{m_{\pi }}{\mu _{\chi }}%
+\varepsilon ^{2}c_{F}(x,\mu )\frac{m_{\pi }^{2}}{\mu _{\chi }^{2}}%
+\varepsilon ^{2}d_{F}(x)\frac{m_{\pi }^{2}}{\mu _{\chi }^{2}}\log \left( 
\frac{m_{\pi }^{2}}{\mu ^{2}}\right) +\mathcal{O}(\varepsilon ^{3})\ , 
\notag \\
\partial _{t}F(x,0,0) &=&\frac{1}{\mu _{\chi }^{2}}\left[ \frac{1}{%
\varepsilon }e_{F}(x)\frac{\mu _{\chi }}{m_{\pi }}+f_{F}(x,\mu
)+g_{F}(x)\log \left( \frac{m_{\pi }^{2}}{\mu ^{2}}\right) +\mathcal{O}%
(\varepsilon )\right] \ ,  \label{F}
\end{eqnarray}%
where all the prefactors are $m_{\pi }$ independent and the $\mu $
dependence of $c_{F}$ and $f_{F}$ cancel the $\mu $ dependence in the
logarithms of the $d_{F}$ and $g_{F}$ terms, respectively.

We will show that those prefactors could have $\delta $ function structures.
Note that instead of using $\delta $ functions with zero widths, one can use
regularized delta functions with finite widths as well. For prefactors with
odd powers of $\varepsilon $ (defined in Eq.(8)), the widths of the delta
functions should be $\sim \varepsilon \sim m_{\pi }/\mu _{\chi }$ or
smaller. For prefactors with even powers of $\varepsilon $, the widths of
the delta functions should be $\sim \varepsilon ^{2}\sim \left( m_{\pi }/\mu
_{\chi }\right) ^{2}$ or smaller. To demonstrate this, we use the
regularized delta function $\delta _{\Delta x}(x)=\left( \Delta x\right)
^{-1}\theta (\Delta x-x)\theta (x)$. The\ $x^{m}$ moment of $\varepsilon
^{n}\delta _{\Delta x}(x)$ gives%
\begin{equation}
\left\langle x^{m}\right\rangle =\int_{0}^{1}dx\varepsilon ^{n}\delta
_{\Delta x}(x)x^{m}=\frac{\varepsilon ^{n}\left( \Delta x\right) ^{m}}{m+1}.
\end{equation}%
When $m>0$, the change of $\left\langle x^{m}\right\rangle $ due to
regularization is $O(\varepsilon ^{n}\left( \Delta x\right) ^{m})$. Thus the
change to the parton distributions due to regularization is $O(\varepsilon
^{n}\Delta x)$. These contributions should be absorbed by matrix elements of
higher order operators in ChPT. However, in ChPT, the higher order operators
are all analytic in light quark mass $m_{q}\left( \propto m_{\pi
}^{2}\right) $. Thus, the regularization is sensible only when $\varepsilon
^{n}\left( \Delta x\right) ^{m}$ is $\varepsilon $ to an even power.

1) For the isovector GPDs $E^{3}$, $E_{T}^{3}$ and $\tilde{H}_{T}^{3}$,
their $\mathcal{O}(\varepsilon )$ contributions are non-vanishing. From Eqs.(%
\ref{aa}), (\ref{xx}) and (\ref{V3}), the $x^{m}$ moment of $E^{3}(x,0,t)$
vanishes except for $m=0:$%
\begin{equation}
\delta E_{m+1}^{3}(0,t)=\varepsilon \delta _{m,0}\mathcal{D}%
_{1,0}^{3,M}\left( t\right) \ .
\end{equation}%
This implies 
\begin{equation}
b_{E^{3}}(x)=\overline{b}_{E^{3}}\delta \left( x\right) \ ,\quad
e_{E^{3}}(x)=\overline{e}_{E^{3}}\delta \left( x\right) \ .
\end{equation}%
where $\overline{b}_{E^{3}}=-12\overline{e}_{E^{3}}=-g_{A}^{2}M/F_{\pi
}=-16.0$. Thus we have%
\begin{eqnarray}
\int d^{2}\mathbf{b}_{\perp }\mathcal{E}^{3}(x,b_{\perp })
&=&a_{E^{3}}(x)+\varepsilon \overline{b}_{E^{3}}\delta \left( x\right) \frac{%
m_{\pi }}{\mu _{\chi }}+\mathcal{O}(\varepsilon ^{2})\ ,  \notag \\
\left\langle b_{\perp }^{2}\right\rangle _{\mathcal{E}^{3}} &=&\frac{4}{%
\varepsilon }\frac{\overline{e}_{E^{3}}\delta \left( x\right) }{\mu _{\chi
}m_{\pi }a_{E^{3}}(x)}+\mathcal{O}(\varepsilon ^{0})\ .  \label{E3}
\end{eqnarray}%
Note that $\int dxa_{E^{3}}(x)$\ is the isovector nucleon anomalous magnetic
moment in the chiral limit. The charge radius square of the isovector
nucleon anomalous magnetic moment is%
\begin{equation}
\left\langle r^{2}\right\rangle _{E^{3}}=\frac{6}{\varepsilon }\frac{%
\overline{e}_{E^{3}}}{\mu _{\chi }m_{\pi }\int dxa_{E^{3}}(x)}+\mathcal{O}%
(\varepsilon ^{0})\ .  \label{101}
\end{equation}

The behavior of $E_{T}^{3}(x,0,t)$ is similar to that of $E^{3}(x,0,t)$ with 
$b_{E_{T}^{3}}(x)\propto \delta \left( x\right) $, $e_{E_{T}^{3}}(x)\propto
\delta \left( x\right) $,$\ $and so on. Also, $\tilde{H}%
_{T}^{3}(x,0,t)=-E_{T}^{3}(x,0,t)/2$ at $\mathcal{O}(\varepsilon )$ from
Eqs. (\ref{aaaa}), (\ref{73}) and (\ref{79}).

2) For the other GPDs, the $\mathcal{O}(\varepsilon )$ contribution to $%
F(x,0,t)$ vanishes such that $b_{F}(x)=e_{F}(x)=0$. The non-analytic
logarithmic terms $d_{F}(x)$ and $g_{F}(x)$ can be predicted in $\chi $PT.
If $g_{F}(x)$ is non-vanishing at certain $x$, then the corresponding $%
\left\langle b_{\perp }^{2}\right\rangle _{\mathcal{F}}$ diverges in the
chiral limit. Here we list the results based on the moment computations.

a) Spin-averaged quark GPDs: The first moment of $H(x,0,0)$ gives the total
number of quarks of certain flavor in the nucleon which is independent of
the quark mass. Thus $\int dxc_{H}(x)=\int dxd_{H}(x)=0$. For the isoscalar
combination $H^{0}$, there are no chiral logarithms at $\mathcal{O}%
(\varepsilon ^{2}),$ 
\begin{equation}
d_{H^{0}}(x)=g_{H^{0}}(x)=0\ .  \label{d0}
\end{equation}%
For the isovector combination $H^{3}$, there are chiral logarithms at $%
\mathcal{O}(\varepsilon ^{2})$ except for the first moment in $x$. This
implies 
\begin{equation}
d_{H^{3}}(x)=-\left( 3g_{A}^{2}+1\right) \left[ a_{H^{3}}(x)-\delta (x)\int
dxa_{H^{3}}(x)\right] \ ,\quad g_{H^{3}}(x)=0\ .
\end{equation}

The isovector GPD $E^{3}$ has been discussed in Eq.(\ref{E3}). The isoscalar
combination $E^{0}+H^{0}$, which satisfies the sum rule in Eq.(\ref{eq1}),
yields 
\begin{equation}
d_{E^{0}}(x)+d_{H^{0}}(x)=-3g_{A}^{2}\left[ a_{E^{0}}(x)+a_{H^{0}}(x)+\,%
\left\langle x\right\rangle _{\pi }\delta ^{\prime }\left( x\right) \right]
\ ,\quad g_{E^{0}}(x)+g_{H^{0}}(x)\propto \delta ^{\prime }\left( x\right) \
,  \label{E0}
\end{equation}%
where $\delta ^{\prime }\left( x\right) =d\delta \left( x\right) /dx$. This
result is derived from Eq.(\ref{r1}) and we have $d_{H^{0}}(x)=0$ from Eq.(%
\ref{d0}).

b) Quark helicity GPDs: $\tilde{E}^{3}$ is not defined at $\xi =0$. For $%
\tilde{E}^{0}$, 
\begin{equation}
d_{\tilde{E}^{0}}(x)=-3g_{A}^{2}a_{\tilde{E}^{0}}(x)-g_{A}^{2}a_{\tilde{H}%
^{0}}(x)\ ,\quad g_{\tilde{E}^{0}}(x)=0\ .
\end{equation}%
And for $\tilde{H}$, 
\begin{equation}
d_{\tilde{H}^{0}}(x)=-3g_{A}^{2}a_{\tilde{H}^{0}}(x)\ ,\quad g_{\tilde{H}%
^{0}}(x)=0\ .
\end{equation}%
and%
\begin{equation}
d_{\tilde{H}^{3}}(x)=-\left( 2g_{A}^{2}+1\right) a_{\tilde{H}^{3}}(x)\
,\quad g_{\tilde{H}^{3}}(x)=0\ .
\end{equation}

c) Quark transversity GPDs: $\tilde{E}_{T}=0$\ when $\xi =0.$\ $\tilde{H}%
_{T}^{3}$\ and $E_{T}^{3}(x,0,t)$\ are already discussed following Eq.(\ref%
{101}) above.

\begin{equation}
d_{H_{T}^{0}}(x)=-3g_{A}^{2}a_{H_{T}^{0}}(x)\ ,\quad g_{H_{T}^{0}}(x)=0\ ,
\end{equation}%
\begin{equation}
d_{H_{T}^{3}}(x)=-\left( 2g_{A}^{2}+\frac{1}{2}\right) a_{H_{T}^{3}}(x)\
,\quad g_{H_{T}^{3}}(x)=0\ ,
\end{equation}%
\begin{equation}
d_{E_{T}^{0}}(x)=-3g_{A}^{2}\left[ a_{E_{T}^{0}}(x)+\frac{4}{3}%
a_{H_{T}^{0}}(x)+2E_{2,0}^{T,\pi ,0}(0)\delta ^{\prime }\left( x\right) %
\right] \ ,\quad g_{E_{T}^{0}}(x)\propto \delta ^{\prime }\left( x\right) \ ,
\label{E0T}
\end{equation}%
\begin{equation}
d_{\tilde{H}_{T}^{0}}(x)=-3g_{A}^{2}\left[ a_{\tilde{H}%
_{T}^{0}}(x)-E_{2,0}^{T,\pi ,0}(0)\delta ^{\prime }\left( x\right) \right]
,\quad g_{\tilde{H}_{T}^{0}}(x)=-\frac{g_{E_{T}^{0}}(x)}{2}\ .  \label{dH0T}
\end{equation}

3) In Ref. \cite{Strikman:2003gz}, Strikman and Weiss argued that the $%
\left\langle b_{\perp }^{2}\right\rangle _{\mathcal{H}_{g}}$ of the gluon
GPD $H_{g}(x,b_{\perp })$ is proportional to $1/m_{\pi }^{2}$ at small $x$ ($%
x\lesssim m_{\pi }/M$) and is proportional to $1/M^{2}$ otherwise. While
this behavior is qualitatively similar to $E^{3}(x,b_{\perp })$ [also $%
E_{T}^{3}(x,b_{\perp })\ $and $\tilde{H}_{T}^{3}(x,b_{\perp })$] described
in Eq.(\ref{E3}), for $H_{g}(x,b_{\perp })$ the result is still inconclusive
in our calculation. We confirm that $\left\langle b_{\perp
}^{2}\right\rangle _{\mathcal{H}_{g}}=\mathcal{O}(1/M^{2})$ or $\mathcal{O}%
(1/\mu _{\chi }^{2})$ when $x\neq 0$ from the result for the isoscalar
moments\ in Eq.(\ref{r1}) but with $m\geq 1$. However, the possible $\delta
(x)$ contribution cannot be probed by the matrix element defined in Eq. (\ref%
{H_g}). We will defer the investigation to a later publication.

Finally, Eqs.(\ref{E0}), (\ref{E0T}) and (\ref{dH0T}) imply that $x\left( 
\mathcal{H}_{g}^{0}+\mathcal{E}_{g}^{0}\right) $, $xE_{Tg}^{0}$ and $x\tilde{%
E}_{Tg}^{0}$ have $\left\langle b_{\perp }^{2}\right\rangle \propto \delta
(x)\log \left( m_{\pi }^{2}\right) $ which diverges in the chiral limit.

\section{\protect\bigskip Conclusions}

Using heavy baryon $\chi $PT we have studied the leading chiral corrections
to the complete set of nucleon GPDs. We have computed the leading quark mass
and momentum transfer dependence of the moments of nucleon GPDs through the
nucleon off-forward twist-2 matrix elements. We have also applied these
results to get insight on the GPDs and their impact parameter space
distributions.

\bigskip

SA and JWC thank the INT at the University of Washington for hospitality.\
SA also thanks the National Taiwan University for hospitality. SA is
supported by Korean Research Foundation and The Korean Federation of Science
and Technology Societies Grant funded by Korean Government (MOEHRD, Basic
Research Promotion Fund). JWC and CWK are supported by the NSC and the NCTS
of ROC.

\end{document}